\documentclass[twocolumn,10pt]{IEEEtran}

\usepackage[utf8]{inputenc}

\usepackage{float}

\usepackage{bm}
\usepackage{amssymb}
\usepackage{amsmath}
\usepackage{graphicx,color}
\usepackage{enumerate}
\usepackage{lipsum} 
\usepackage[capitalise]{cleveref} 
\usepackage{tikz}
\usetikzlibrary{patterns}
\usepackage{accents}

\usepackage{enumerate}
\usepackage{comment}
\usepackage{bm}
\usepackage{algorithm}
\usepackage{algorithmic}

\usepackage{cite}

\usepackage{soul}
\usepackage[normalem]{ulem} 

\usepackage[caption=false]{subfig}

\graphicspath{{figs/}}

\DeclareMathOperator{\trace}{trace}

\newtheorem{proposition}{Proposition}
\newtheorem{remark}{Remark}
\newtheorem{theorem}{Theorem}
\newtheorem{definition}{Definition}

\newtheorem{lemma}{Lemma}
\newtheorem{example}{Example}

\usepackage{url}

\newcommand{\mmv}[1]{{\color{magenta}\bf [#1]}}
\newcommand{\xz}[1]{{\color{blue} #1}}

\DeclareMathOperator{\Equaldef}{\overset{def}{=}}

\DeclareMathOperator{\indep}{\perp \!\!\! \perp}

\renewcommand{\Pr}{\mathbf{P}}

\title{Robust one-shot estimation over shared networks in the presence of denial-of-service attacks}

\author{Xu Zhang and Marcos M. Vasconcelos
\thanks{X. Zhang was partially supported by China National Postdoctoral Program for Innovative Talents (No. BX2021346), China Postdoctoral Science Foundation (No. 2022M713316), and National Natural Science Foundation of China (No. 12288201). M. M. Vasconcelos was partially supported by the Commonwealth Cyber Initiative.}
\thanks{X.~Zhang is with LSEC, Academy of Mathematics and Systems Science, Chinese Academy of Sciences, Beijing, China (e-mail: \texttt{xuzhang\_cas@lsec.cc.ac.cn}).}
\thanks{M. M. Vasconcelos is  with the Department of Electrical Engineering at the FAMU-FSU College of Engineering, Florida State University, USA (e-mail: \texttt{mm22eo@fsu.edu}).}

}
\date{}

\allowdisplaybreaks[4]

\begin{document}

\maketitle

\begin{abstract}
Networked multi-agent systems often communicate information over low-power shared wireless networks in unlicensed spectrum, which are prone to denial-of-service attacks. An instance of this scenario is considered: multiple pairs of agents, each pair consisting of a transmitting sensor and a receiver acting as an estimator, communicate strategically over shared communication networks in the presence of a jammer who may launch a denial-of-service attack in the form of packet collisions. Using the so-called \textit{coordinator approach}, we cast this problem as a zero-sum Bayesian game between the coordinator, who jointly optimizes the transmission and estimation policies, and a jammer who optimizes its probability of performing an attack. We consider two cases: point-to-point channels and large-scale networks with a countably infinite number of sensor-receiver pairs. When the jammer proactively attacks the channel, we find that this game is nonconvex from the coordinator's perspective. However, we construct a saddle point equilibrium solution for any multi-variate Gaussian input distribution for the observations despite the lack of convexity. In the case where the jammer is reactive, we obtain a customized algorithm based on sequential convex optimization, which converges swiftly to first order Nash-equilibria. Interestingly, we discovered that when the jammer is reactive, it is often optimal to block the channel even when it knows that the channel is idle to create ambiguity at the receiver. 
\end{abstract}
\section{Introduction}

Networked multi-agent systems often consist of multiple decision making agents collaborating to perform a task. Popular examples of network systems include ground and/or aerial robotic networks, sensor networks, and the internet of things. In order to achieve a synergistic behavior, the agents often communicate messages over a wireless network among themselves. Typically, a network system architecture will also involve one or multiple nodes communicating with a gateway or base-station. Over these links, the transmitting agent sends messages containing one or more state variables that need to be estimated at the base-station. In particular, remote sensing where one (or multiple) sensor(s) communicates its measurements over a shared wireless channel to one or more non-collocated access points or base-stations is a fundamental building block of many cyber-physical systems \cite[and references therein]{Vasconcelos:2020,xia2016networked,Kang:2023}. 

There are many communication protocols that enable such local communication such as Bluetooth, wi-fi and cellular, among others. The choice of a given protocol requires meeting some specifications, but there is no single protocol that achieves all desirable characteristics and uniformly better than the others. For example, Low Power Wide Area Networks (LPWANs) provide power efficiency and large coverage leading to very cost efficient deployments \cite{Raza2017low}. However, such protocols operate in frequency bands in the so-called \textit{unlicensed spectrum}, and are therefore vulnerable to malicious agents interested in disrupting the communication link between the anchor-node(s) and the base station using denial-of-service attacks. Denial-of-Service (DoS) is a class of cyber-attacks where a malicious agent, often referred to as the \textit{jammer}, may disrupt the communication link between the legitimate transmitter-receiver pair. DoS attacks are widely studied at different levels of modeling detail of the communication channel. For example if the channel is assumed to be a physical layer model, the jammer may introduce additional Gaussian noise to the transmitted signal. If the channel is modeled at the network layer by a packet-drop channel, the jammer may increase the probability of dropping a packet. We consider a medium access control (MAC) layer model in which the jammer may decide to block the channel by transmitting an interference signal that overwhelms the receiver, causing a packet collision. 



\begin{figure}[t!]
    \centering
    \includegraphics[width=0.95\columnwidth]{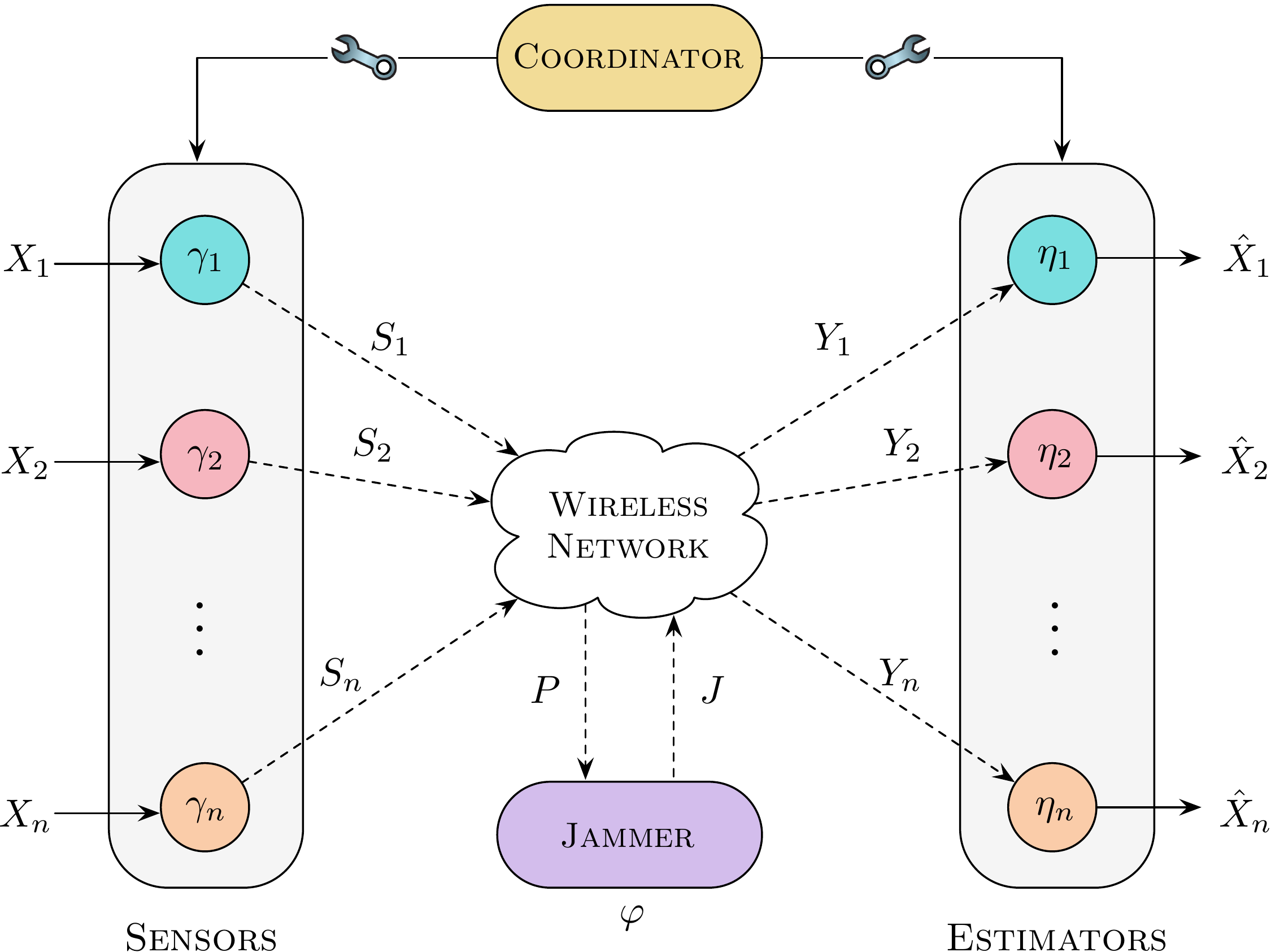}
    \caption{Block diagram for a remote estimation game between a coordinator and a jammer. The jammer may have access to side information on the channel's occupancy. The coordinator designs the policies for the sensor and the estimator.}
    \label{fig:system}
\end{figure}

We consider the remote estimation system depicted in \cref{fig:system}, which is comprised of multiple sensor and estimator pairs communicating over a shared wireless network modeled by a collision channel in the presence of a jammer.
Each sensor makes a stochastic measurement $X_i$ of a physical quantity according to a given distribution, and decides whether to transmit it or not to the corresponding estimator. Communication is costly, therefore, the sensors must transmit wisely. We consider two cases: 1. the \textit{proactive jammer} that cannot sense if the channel is being used by the sensors; 2. the \textit{reacitve jammer} that can sense the channel, i.e., has access to the number of transmitting sensors $P$. 
Jamming is assumed to be costly, therefore, the jammer must act strategically. 

Finally, each estimator observes the channel output and declares an corresponding estimate $\hat{X}_i$ for the sensor's observation such as to minimize the expected quadratic distortion between $X_i$ and $\hat{X}_i$. We study this problem as a zero-sum game between a coordinator (system designer) and the jammer. Our goal is to characterize equilibrium solutions and obtain efficient algorithms to compute them. The main difference between our model and existing work in this area is the presence of a virtual binary signaling channel that can be exploited by the coordinator to guarantee a minimum level of performance of the system  in the presence of DoS attacks. 


There exists an extensive literature on strategic communication in the presence of jammers. This class of problems seems to have started with the seminal work of Basar \cite{basar1983gaussian}, which obtained a complete characterization of the saddle point equilibria when the sensor measurements and the channel are Gaussian. Recently, an extension to the two-way additive Gaussian noise channel was studied by McDonald et al. in \cite{mcdonald2019two}. A jamming problem where the transmitter and estimator have different objectives was solved by Akyol et al. in \cite{Akyol2017Info} using a hierarchical game approach. A jamming problem with and without common randomness between the transmitter and estimator is studied Akyol in \cite{Akyol2019Optimal} and a Stackelberg game formulation was considered by Gao et al. \cite{Gao2019Communication}. Another interesting problem formulation is due to Shafiee and Ulukus in \cite{Shafiee2009Mutual}, where the pay-off function is the mutual information between the channel input and output. Jamming over fading channels was considered by Ray et al. in \cite{ray2006optimal} and subsequently by Altman et al. in \cite{Altman:2011}. An LTE network model was considered by Aziz et al. in \cite{aziz2020resilience}.

Another class of remote estimation problems focuses on the state estimation of a linear time invariant system driven by Gaussian noise under DoS attacks. Li et al. \cite{li2015jamming} studied a jamming game where the transmitter and jammer have binary actions. A SINR-based model was considered by Li et al. in \cite{li2016sinr}, where the transmitter and jammer decide among multiple discrete power levels. The case of continuum of power levels was studied by Ding et al. in \cite{ding2017stochastic}. \textcolor{black}{A jamming model over a channel with two modes (i.e., free mode and safe mode) was analyzed by Wu et al. in \cite{Wu:2017}.}
A jamming problem with asymmetric feedback information and multi-channel transmissions was considered by Ding et al. in \cite{ding2018attacks} and \cite{ding2017multi}, respectively. A Stackelberg equilibrium approach to this problem was considered by Feng et al. in \cite{Feng:2021}. The problem of optimizing the attack scheduling policy from the jammer's perspective was considered by Peng et al. in \cite{peng2017optimal}.  

The model described herein is closely related to the work of Gupta et al. \cite{gupta2012dynamic,gupta2016} and Vasconcelos and Martins \cite{vasconcelos:2017a,vasconcelos2018optimal}, where there is a clear distinction between the channel being blocked vs. idle. As in \cite{gupta2012dynamic}, we assume that
the transmission decision $U$ may be available as side information to the jammer, but not the full input signal $X$. This assumption is realistic in the sense that the bits used to encode $X$ may be encrypted. In the game considered in \cite{gupta2012dynamic}, it is assumed that the receiver is fixed, and the game is played between the sensor and the jammer. Instead, we follow Akyol \cite{Akyol2019Optimal} in which the sensor and estimator are distinct agents implementing policies optimized by a \textit{coordinator} \cite{Nayyar2014}\footnote{A subset of the results for the point-to-point channel reported herein have appeared in \cite{zhang2022robust}, which uses a different optimization technique based on so-called \textit{rearrangement inequalities}. In the present paper, we used different optimization techniques that allow a complete generalization for the multivariate Gaussian observation model. Such generalization would not be possible using the same techniques in \cite{zhang2022robust}. Moreover, this paper introduces the analysis for the large-scale case, which appears here for the first time and does not follow from the analysis in \cite{zhang2022robust}.}. 

{\color{black}
The main contributions of the paper are summarized as follows:
\begin{enumerate}
\item For the proactive jammer over a point-to-point channel, we provide the optimal strategies for the coordinator and the jammer that constitute a saddle point equilibrium, which appears as two scenarios depending on the transmission and jamming costs. This result holds even though the objective function is non-convex from the coordinator’s perspective. 
\item For the reactive jammer over a point-to-point channel, we propose alternating between Projected Gradient Ascent (PGA) and Convex-Concave Procedure (CCP) to achieve an approximate first-order Nash-equilibrium. Our numerical results demonstrate that the proposed PGA-CCP algorithm exhibits superior convergence rates compared to the traditional Gradient Descent Ascent (GDA) algorithm. A significant contribution here is that the optimal estimator employs representation symbols with distinct values for the no-transmission and collisions, as opposed to when the jammer is proactive, which uses the mean to estimate the observations in both situations.
\item For large-scale networks we assume a problem with a countably infinite number of agents \cite[and references therein]{Sanjari:2021} under the possibility of a proactive jamming attack. We compute the limiting objective function when the normalized channel capacity converges to a constant $\bar{\kappa}$. In this regime, the zero-sum game between the coordinator and the jammer over large-scale networks is equivalent to a constrained minimax problem. We establish the saddle point equilibrium of the optimal strategies for the coordinator and the jammer, which consists of six scenarios based on transmission cost, jamming cost, and the normalized capacity.
\end{enumerate}

}
\section{System Model}

Consider a remote sensing system consisting of $n$ sensors. Let $[n]$ denote the set $\{1,\cdots,n\}$. Each sensor makes a random measurement, which is represented by a random vector. Let $X_i \in \mathbb{R}^{m}$ denote the measurement of the $i$-th sensor. We assume for tractability that the measurements are independent and identically distributed Gaussian random vectors across sensors, that is, $X_i \sim \mathcal{N}(\mu,\Sigma),$ $i\in[n]$.

We denote the probability density function (pdf) of a multivariate Gaussian random vector by:
\begin{equation}
f(x) \Equaldef \frac{1}{\sqrt{(2\pi)^m|\Sigma|}}\exp\Big(-\frac{(x-\mu)^\mathsf{T}\Sigma^{-1}(x-\mu)}{2}\Big).
\end{equation}


The goal of the sensors is to communicate their measurements to one or multiple receiver over a shared wireless network of limited capacity in the presence of a jammer. 

\subsection{Transmitters}
We define the following collection of policies: 
\begin{equation}
    \gamma=(\gamma_1,\cdots,\gamma_n).
\end{equation}

\begin{definition}[Transmission policy]
A transmission policy for the $i$-th sensor is a measurable function $\gamma_i: \mathbb{R}^{m}\rightarrow [0,1]$ such that
\begin{equation}
    \Pr(U_i=1 \mid X_i=x_i) = \gamma_i(x_i), \ \ i \in [n].
\end{equation}
\end{definition}

When the $i$-th sensor makes a transmission, it sends a packet containing its identification number and its observed measurement as follows. Given $X_i=x_i$ and $U_i=1$, the signal transmitted to the receiver is:
\begin{equation}
    S_i = (i,x_i).
\end{equation}

The reason this is done is to remove the ambiguity regarding the origin of each measurement, since they could correspond to physical quantities captured at different locations, or, potentially, completely different physical quantities.

When a sensor does not transmit, we assume that the signal transmitted corresponds to an \textit{empty} packet, which is mathematically represented by 
\begin{equation}
    S_i = \varnothing.
\end{equation}

When a sensor transmits, it encodes the data using a cryptographic protocol. Typically, the LoRaWAN IoT standard uses the 128bit-AES lightweight encryption. The nature of the protocol is not important here, but it implies that when the attacker \textit{senses} the channel, it cannot decode the content in each transmitted packet. However, it is capable of detecting whether a given channel is used based on a threshold detector on the power level in the channel's frequency band. 

We assume that the communication occurs via a wireless medium of capacity $\kappa(n) \in (0,n)$. Notice that the capacity of the channel corresponds to the number of packets that the channel can support simultaneously, and is not related to the information theoretic notion of capacity.

Provided that the channel is not blocked by the attacker, when the total number of transmitting sensors is below or equal to the channel capacity, the receiver observes the packets perfectly. Conversely, when the number of simultaneous transmissions exceeds the channel capacity, the receiver observes a \textit{collision symbol}.
We represent this as follows:
Let
\begin{equation}
    P \Equaldef \sum_{i=1}^nU_i
\end{equation}
and
\begin{equation}
    Y = \begin{cases}
    \{S_i\}_{i=1}^n & \text{if} \ \ P \leq \kappa(n) \\ 
    \mathfrak{C} & \text{otherwise.}
    \end{cases}
\end{equation}

One feature of the wireless medium is that it is prone to malicious denial of service attacks known as jamming. There are many types of jamming attacks, but here we focus on two kinds: the \textit{proactive} and the \textit{reactive} jammer. 

\subsection{Proactive jamming}
We define the \textit{proactive} jammer, as one that decides whether to block the channel or not without sensing the channel. Therefore, at each time instant, the decision to attack is made according to a mixed strategy, such that with a certain probability, it spends a fixed amount of energy to block the network. At the receiver, the jamming attack is perceived as if a collision among many packets has happened. 

For the proactive jammer, the decision to block the channel or not is denoted by the variable $J$, which is independent of $P$, i.e.,
\begin{equation}
    \Pr(J=1) = \varphi \in [0,1].
\end{equation}

\subsection{Reactive jamming}
{\color{black}
Reactive jamming is a more sophisticated attack model in which the jammer first senses whether the channel is occupied or not. Then the jammer adjusts its probability of blocking the channel based on the channel state. 
The reactive jamming strategy is characterized by a vector $\varphi\Equaldef (\alpha,\beta) \in [0,1]^2$:
\begin{align}
	\Pr\big(J=1 \mid U=0 \big) =\alpha, ~~ \Pr\big(J=1 \mid U=1\big) =\beta,
\end{align}
where $\alpha$ is the jamming probability when the channel is not occupied and $\beta$ is the jamming probability when the channel is occupied.


}

\subsection{Channel output}

Given $\{X_i=x_i\}_{i=1}^n$, let us define the channel output alphabet as the $\mathcal{Y}(x_1,\cdots,x_n)$ as follows. Let  
\begin{equation}
   \mathcal{Y}_i(x_i) \Equaldef \big\{ \varnothing, (i,x_i)\big\}. 
\end{equation}
Then, 
\begin{equation}
   \mathcal{Y}(x_1,\cdots,x_n) \Equaldef \big(\mathcal{Y}_1(x_1) \times \cdots \times \mathcal{Y}_n(x_n) \big) \cup \mathfrak{C}. 
\end{equation}

The channel output is given by:
\begin{equation}
    Y = \begin{cases}
    \{S_i\}_{i=1}^n & \text{if} \ \ P\leq \kappa(n),\ J = 0 \\ 
    \mathfrak{C} & \text{otherwise.}
    \end{cases}
\end{equation}

\subsection{Receiver}
Finally, we define the receiver's policy. Let the receiver policy $\eta$ be a collection of functions
\begin{equation}
    \eta=(\eta_1,\cdots,\eta_n),
\end{equation}
where $\eta_i:\mathcal{Y}_i(x_i)\cup \mathfrak{C} \rightarrow \mathbb{R}^{m}$ is a measurable map, $i\in[n]$. Assume that from $Y$, the receiver forms $n$ signals $\{Y_i\}_{i=1}^n$ such that
\begin{equation}
Y_i = \begin{cases}
\mathfrak{C}, & \text{if} \ \ Y = \mathfrak{C} \\
S_i, & \text{otherwise}.
\end{cases}  
\end{equation}

Given $X_i=x_i$, an estimation policy is a measurable map $\eta_i : \mathcal{Y}(x_i) \rightarrow \mathbb{R}^m$ such that 
\begin{equation}
    \hat{X}_i = \eta_i(Y_i), \ \ i\in[n].
\end{equation}



We assume that a coordinator plays the role of a system designer, and jointly adjusts the transmission and estimation policies at the sensors and the receiver in the presence of a jammer. The approach is akin to a \textit{robust design} problem in which, the coordinator seeks to optimize the performance of the distributed sensing system, when the operation may be affected by a DoS attack. We assume that the coordinator \textit{plays a zero-sum game} with the attacker, where the objective function is given by:
\begin{multline} \label{eq:objectiven}
    \mathcal{J}_n\big((\gamma,\eta),\varphi \big) \Equaldef \frac{1}{n}\mathbf{E}\Bigg[\sum_{i=1}^n \big[ \|X_i-\hat{X}_i\|^2 +c\mathbf{1}(U_i=1)\big]\Bigg] \\ - d\mathbf{P}(J=1).
\end{multline}
Note that even when there are multiple sensors, there are only two players, namely, the coordinator and the jammer. 

We are interested in obtaining policies that constitute saddle point equilibria.

\vspace{5pt}

\begin{definition}[Saddle point equilibrium] A policy tuple $(\gamma^\star,\eta^\star,\varphi^\star)$  is a saddle point equilibrium if
\begin{equation}\label{eq:SPE}
\mathcal{J}\big( (\gamma^\star,\eta^\star),\varphi \big) \leq \mathcal{J}\big( (\gamma^\star,\eta^\star),\varphi^\star \big) \leq \mathcal{J}\big( (\gamma,\eta),\varphi^\star \big),
\end{equation}
for all $\gamma,\eta,$ and $\varphi$ in their respective admissible policy spaces.
\end{definition}



%


\section{Point-to-point channels}

We start our analysis by considering a point-to-point channel
that can support at most one packet per time-slot, i.e., $n=1$  and $\kappa(n)=1$. In this case, $P=U_1$, the objective function becomes 
\begin{equation}\label{eq:objective_n_1}
		\mathcal{J}_1\big( (\gamma,\eta),\varphi \big) = \mathbf{E}\big[ \|X_1-\hat{X}_1\|^2 \big] + c\Pr(U_1=1) - d\Pr(J=1).
\end{equation}
From here on, we will ignore the subscripts to simplify the notation. The first step is to assume that, without loss of generality, the estimator at the receiver implements the following map\footnote{Here we omit that the received signal includes the identification number. This information is redundant in the point-to-point setting.}
\begin{equation} \label{eq:estimator1}
	\eta(y) =\begin{cases}
		x & \text{if} \ \ y=x \\ 
		\hat{x}_0 & \text{if}\ \ y=\varnothing\\ 
		\hat{x}_1 & \text{if}\ \ y=\mathfrak{C},
	\end{cases}
\end{equation}
where the variables $\hat{x}_0$ and $\hat{x}_1$ serve as representation symbols for the no-transmission and collision events, and will be optimized by the coordinator. Let 
$\hat{x} \Equaldef  (\hat{x}_0,\hat{x}_1).
$
The estimation policy in \cref{eq:estimator1} is parametrized by $\hat{x} \in \mathbb{R}^{2m}$.

\subsection{Proactive jamming of point-to-point collision channels}\label{sec:pt2pt}
We obtain the following structural result for the set of optimal transmission policies at the sensor.

\vspace{5pt}

\begin{proposition}[Optimality of threshold policies]\label{prop:threshold_n_1} For a point-to-point system with a proactive attacker with a fixed jamming probability $\varphi \in[0,1]$, and an arbitrary estimation policy $\eta$ indexed by representation symbols $\hat{x}\in\mathbb{R}^{2m}$, the optimal transmission strategy is\footnote{The function $\mathbf{1}(\mathfrak{S})$ denotes the indicator function of the Boolean statement $\mathfrak{S}$, i.e., $\mathbf{1}(\mathfrak{S})=1$ if $\mathfrak{S}$ is true, and $\mathbf{1}(\mathfrak{S})=0$ if $\mathfrak{S}$ is false.}:
	\begin{equation}
		\gamma_{\eta,\varphi}^\star(x) = 
		\mathbf{1}\big( (1-\varphi)\|x-\hat{x}_0\|^2 > c \big).
	\end{equation}
\end{proposition}

\vspace{5pt}



\begin{IEEEproof} 
	Using the law of total expectation, the definition of the estimation policy in \cref{eq:estimator1}, and the fact that $(U,X) \indep J$, we rewrite \cref{eq:objective_n_1} as follows:
	\begin{multline} \label{eq:intermediate}
		\mathcal{J}\big( (\gamma,\eta),\varphi \big) = 
		\mathbf{E}\big[\|X-\hat{x}_0\|^2 \mid U=0 \big]\Pr(U=0)(1-\varphi) \\ 
		+ \mathbf{E}\big[\|X-\hat{x}_1\|^2 \big]\varphi + c\Pr(U=1) -d\varphi,  
	\end{multline}
	where $\varphi =\Pr(J=1)$. 
	\Cref{eq:intermediate} is equivalent to
	\begin{multline}
		\mathcal{J}\big( (\gamma,\eta),\varphi \big) = \int_{\mathbb{R}^{m}} (1-\varphi)\|x-\hat{x}_0\|^2\big(1-\gamma(x)\big)f(x)\mathrm{d} x \\
		+ \int_{\mathbb{R}^{m}}c\gamma(x)f(x)\mathrm{d} x + \varphi \mathbf{E}\big[\|X-\hat{x}_1\|^2 \big] -d\varphi.
	\end{multline}
	
	Finally, when optimizing over $\gamma$ for fixed $\eta$ and $\varphi$, we have an infinite dimensional linear program with the following constraint:
	\begin{equation}
		0 \leq \gamma(x) \leq 1,  \ \ x\in\mathbb{R}^{m}.
	\end{equation}
	The solution to this problem is obtained by comparing the arguments of the two integrals that involve $\gamma$, i.e., $x \in \{\xi \mid \gamma^\star_{\eta,\varphi}(\xi) = 1\}$ if and only if $(1-\varphi)\|x-\hat{x}_0\|^2 > c.$
\end{IEEEproof}

\vspace{5pt}

\begin{remark}
\Cref{prop:threshold_n_1} implies that the optimal transmission policy is always of the threshold type. Moreover, this threshold policy is symmetric if and only if $\hat{x}_0=0$. The optimal policy is non-degenerate if $\varphi \in[0,1)$, or degenerate when $\varphi=1$. The latter corresponds to a \textit{never-transmit} policy.
\end{remark}

\vspace{5pt}

With a slight abuse of notation, the structure of the optimal transmission policy in \cref{prop:threshold_n_1} implies that the objective function \cref{eq:objective_n_1} assumes the following form:
\begin{multline}\label{eq:new_objective_n_1}
\mathcal{J}\big((\gamma_{\eta,\varphi}^\star,\eta),\varphi\big) = \mathbf{E}\bigg[\min\Big\{(1-\varphi)\|X-\hat{x}_0\|^2, c \Big\} \bigg] \\ + \varphi\Big(\mathbf{E}\big[\|X-\hat{x}_1\|^2 \big] -d \Big) \Equaldef \tilde{\mathcal{J}}(\hat{x},\varphi).
\end{multline}

\vspace{5pt}

\begin{proposition}\label{prop:properties}
Let $X\in \mathbb{R}^m$ be a Gaussian random vector with mean $\mu$ and covariance $\Sigma$. The function $\tilde{\mathcal{J}}(\hat{x},\varphi)$ is non-convex in $\hat{x}\in\mathbb{R}^{2m}$ and concave in $\varphi \in [0,1]$.
\end{proposition}

\vspace{5pt}

\begin{IEEEproof} \textbf{Non-convexity in} $\boldsymbol{\hat{x}}$ -- We set $\varphi=0.5$, $c=d=1$ and $X \sim \mathcal{N}(0,1)$. We can numericall verify that:
\begin{equation}
\frac{1}{2}\tilde{\mathcal{J}}\big((0,0),\varphi\big)+
\frac{1}{2}\tilde{\mathcal{J}}\big((1,0),\varphi\big) < \tilde{\mathcal{J}}\big((0.5,0),\varphi\big).
\end{equation}

\noindent \textbf{Concavity in $\boldsymbol{\varphi}$} --  Define $p:\mathbb{R}^{m}\times\mathbb{R}^{m}\times\mathbb{R}\rightarrow \mathbb{R}$ such that 
\begin{equation}
	p(x,\hat{x}_0,\varphi)\Equaldef \min\Big\{(1-\varphi)\|x-\hat{x}_0\|^2, c \Big\}.
\end{equation} 
For fixed $x, \hat{x}_0 \in \mathbb{R}^{m}$, $p(x,\hat{x}_0,\varphi)$ is the pointwise minimum of affine functions in $\varphi$. Therefore, it is concave for all $x\in\mathbb{R}^{m}$. Taking the expectation of $p(X,\hat{x}_0,\varphi)$ with respect to $X$ preserves the concavity in $\varphi$. 


\end{IEEEproof}

We proceed by minimizing \cref{eq:new_objective_n_1} with respect to the estimation policy, which is a non-convex finite dimensional optimization problem over $\hat{x}\in\mathbb{R}^{2m}$. A classic result in probability theory implies that $\hat{x}_1^\star = \mu$. However, due to the lack of convexity it is non-trivial to find the minimizer $\hat{x}_0^\star$ for an arbitrary Gaussian distribution.

\subsection{The scalar case}

We begin with the result for the scalar case. The general vector case is discussed in Appendix \ref{appendix: p2p_vector}.

\vspace{5pt}

\begin{theorem}[Optimal estimator for scalar Gaussian sources] \label{thm: optimal_estimator_n_1} Let $X$ be a Gaussian random variable with mean $\mu$ and variance $\sigma^2$. The optimal estimator is
\begin{equation} 
	\eta^\star (y) = \begin{cases}
		\mu, & \text{if} \ \ y \in \{\varnothing,\mathfrak{C}\} \\
		x, & \text{if} \ \ y = x.
	\end{cases}
\end{equation}
\end{theorem}

\begin{IEEEproof} Since $\hat{x}_1^\star = \mu$, after ignoring the constants, 
the objective function becomes 
\begin{multline}
	\tilde{\mathcal{J}}(\hat{x},\varphi) = \int_{-\infty}^{+\infty} \min\Big\{(1-\varphi)(x-\hat{x}_0)^2, c \Big\} \frac{e^{-\frac{(x-\mu)^2}{2\sigma^2}}}{\sqrt{2\pi \sigma^2}}  \mathrm{d} x.
\end{multline}

After a change of variables, the objective function may be expressed as
\begin{multline}
	\tilde{\mathcal{J}}(\hat{x},\varphi) 
	= \int_{-\infty}^{+\infty} \min\Big\{(1-\varphi)z^2, c \Big\} \frac{e^{-\frac{(z+\hat{x}_{0}-\mu)^2}{2\sigma^2}}}{\sqrt{2\pi \sigma^2}}  \mathrm{d} z.
\end{multline}	

Taking the partial gradient of $\tilde{\mathcal{J}}(\hat{x},\varphi)$ with respect to $\hat{x}_0$ we obtain 
\begin{IEEEeqnarray}{rCl}
	\nabla_{\hat{x}_0}\tilde{\mathcal{J}}(\hat{x},\varphi)  & = & -\int_{-\infty}^{+\infty}  \min\Big\{(1-\varphi)z^2, c \Big\} \frac{e^{-\frac{(z+\hat{x}_{0}-\mu)^2}{2\sigma^2}}}{\sqrt{2\pi \sigma^2}}  \nonumber \\
	&  & \qquad \qquad \cdot \left(\frac{z+\hat{x}_{0}-\mu}{\sigma^2}\right)\mathrm{d} z
	\\ & \stackrel{(a)}{=} & - \int_{-\infty}^{+\infty}  \min\Big\{(1-\varphi)[v-(\hat{x}_0-\mu)]^2, c \Big\}   \nonumber\\
	& & \qquad \qquad \cdot \frac{e^{-\frac{v^2}{2\sigma^2}}}{\sqrt{2\pi \sigma^2}}\left(\frac{v}{\sigma^2}\right)\mathrm{d} v,
\end{IEEEeqnarray}	
where $(a)$ follows by exchanging $z+\hat{x}_{0}-\mu$ with $v$. Let 
\begin{align}
	h(v)&\Equaldef \min\Big\{(1-\varphi)[v-(\hat{x}_0-\mu)]^2, c \Big\},\\
	g(v)& \Equaldef\frac{1}{\sqrt{2\pi \sigma^2}} e^{-\frac{v^2}{2\sigma^2}}\cdot \Big(-\frac{v}{\sigma^2}\Big).
\end{align}
Note that $g(v)$ is an odd function with $g(v)<0$ for $v>0$ and $h(v)$ is nonnegative for all $v$. We analyze the sign of $\nabla_{\hat{x}_0}\tilde{\mathcal{J}}(\hat{x},\varphi)$ in three cases:
\begin{enumerate}
\item For $\hat{x}_0=\mu$, $h(v)$ is an even function, which implies that $h(v)g(v)$ is an odd function. Therefore,
\begin{equation}
\nabla_{\hat{x}_0}\tilde{\mathcal{J}}(\hat{x},\varphi)=\int_{-\infty}^{+\infty} h(v)g(v)\mathrm{d} v=0.
\end{equation}
\item For $\hat{x}_0>\mu$, we have $0 \le h(v) < h(-v)$ when $v\ge 0$. Since  $g(v)$ is an odd function and $g(v)<0$ for $v>0$, we have
\begin{equation}
	 \nabla_{\hat{x}_0}\tilde{\mathcal{J}}(\hat{x},\varphi)
	=\int_{0}^{+\infty} \big(h(v)-h(-v)\big)g(v)\mathrm{d} v>0.	
\end{equation}
\item For $\hat{x}_0<\mu$, we have $0 \le h(-v) < h(v)$ when $v\ge 0$.  Since $g(v)$ is an odd function and $g(v)<0$ for $v>0$, we have
\begin{equation}
	\nabla_{\hat{x}_0}\tilde{\mathcal{J}}(\hat{x},\varphi) =\int_{0}^{+\infty} \big(h(v)-h(-v)\big)g(v)\mathrm{d} v<0.	
\end{equation}
\end{enumerate}

Therefore, we conclude that $\hat{x}_0=\mu$ is the unique minimizer of $\tilde{\mathcal{J}}(\hat{x},\varphi)$.
\end{IEEEproof}

\vspace{5pt}

Without loss of generality, for the remainder of this section we assume that $\mu=0$. 
The optimal transmitter and estimator's strategies for a symmetric Gaussian distribution imply that the objective function for the jammer is given by
\begin{equation}\label{eq:jammer_obj}
	\mathcal{J}\big((\gamma_{\eta^\star,\varphi}^\star,\eta^\star),\varphi\big) = \mathbf{E}\bigg[\min\Big\{(1-\varphi) X^2, c \Big\} \bigg]  + \varphi\Big(\mathbf{E}\big[X^2 \big] -d \Big).
\end{equation}

From \cref{prop:properties}, the objective function in \cref{eq:jammer_obj} is concave with respect to $\varphi$. Therefore, 
we can compute the optimal jamming probability $\varphi^\star$. 
Let $\tilde{\varphi}$ be defined as 
\begin{equation} \label{eq: J_d_beta_tilde}
\tilde{\varphi} \Equaldef \inf \bigg\{ \varphi \in [0,1) \ \Big| \  \int_{\sqrt{c/(1-\tilde{\varphi})}}^{+\infty} x^2f(x)\mathrm{d} x =\frac{d}{2}\bigg\}.
\end{equation}

\vspace{5pt}

{ \color{black}
\begin{theorem}[Optimal jamming probability for scalar Gaussian sources] \label{thm: optimal_jam_prob} 
Let $X$ be a Gaussian random variable with mean $0$ and variance $\sigma^2$. 
The optimal jamming probability for the optimal transmission policy in \cref{prop:threshold_n_1} and  the optimal estimation policy in \cref{thm: optimal_estimator_n_1} is
	\begin{equation}
		\varphi^\star=
		\left\{\begin{array}{ll}
			 {\tilde{\varphi}} & {\text { if }  \int_{\sqrt{c}}^{+\infty} x^2f(x)\mathrm{d} x \ge d/2} \\
			 {0} & {\text { otherwise. }} 
		\end{array}\right.
	\end{equation}
\end{theorem}
\vspace{5pt}
\begin{IEEEproof}
	First, we represent \cref{eq:jammer_obj} in integral form as
	\begin{multline}
		\mathcal{J}\big((\gamma_{\eta^\star,\varphi}^\star,\eta^\star),\varphi\big) = \int_{-\sqrt{c/(1-\varphi)}}^{\sqrt{c/(1-\varphi)}} (1-\varphi)x^2f(x)\mathrm{d} x \\ + 2 \int_{\sqrt{c/(1-\varphi)}}^{+\infty} c f(x) \mathrm{d}x 
		+\varphi \Big(\mathbf{E}\big[X^2 \big] -d \Big).
	\end{multline}
	Taking the derivative of  the objective function with respect to $\varphi$, we have
	\begin{IEEEeqnarray}{rCl} \label{eq: J_d_beta}
		\mathcal{G}(\varphi) & \Equaldef & {\nabla_\varphi}\mathcal{J}\big((\gamma_{\eta^\star,\varphi}^\star,\eta^\star),\varphi\big)  \\
		& = & 2\int_{\sqrt{c/(1-\varphi)}}^{+\infty} x^2f(x)\mathrm{d} x -d.
	\end{IEEEeqnarray}
	
	Notice that $\mathcal{G}(\varphi)$ is a monotone decreasing function with respect to $\varphi$ and the following identities hold
	\begin{align}
		&\mathcal{G}(0)  =   2\int_{\sqrt{c}}^{+\infty} x^2f(x)\mathrm{d} x -d \\
		&\lim_{\varphi\uparrow 1}\mathcal{G}(\varphi)  =   -d.
	\end{align}
	
	If $\mathcal{G}(0)\ge0$, then the optimal $\varphi^\star = \tilde{\varphi}$ due to the fact that $\mathcal{G}(\tilde{\varphi})= 0$.	
	If $\mathcal{G}(0)<0$, the objective function is decreasing in $\varphi$. Therefore, $\varphi^\star=0$.
	
\end{IEEEproof}

\vspace{5pt}

\begin{lemma} 	\label{col:LHS_saddle}
	Let $X$ be a Gaussian random variable with mean $0$ and variance $\sigma^2$. The optimal jamming probability $\varphi^\star$ satisfies
	\begin{equation}
		\mathcal{J}\big( (\gamma^\star_{\eta^\star,\varphi^\star},\eta^\star,\varphi \big) \leq \mathcal{J}\big( (\gamma^\star_{\eta^\star,\varphi^\star},\eta^\star),\varphi^\star \big).
	\end{equation}
\end{lemma}

\vspace{5pt}

\begin{IEEEproof} Consider the objective function in integral form as
	\begin{multline}
		\mathcal{J}\big((\gamma_{\eta^\star,\varphi^\star}^\star,\eta^\star),\varphi\big) = \int_{-\sqrt{c/(1-\varphi^\star)}}^{\sqrt{c/(1-\varphi^\star)}} x^2f(x)\mathrm{d} x \\ + 2 \int_{\sqrt{c/(1-\varphi^\star)}}^{+\infty} c f(x) \mathrm{d}x 
		+\varphi \Big(2\int_{\sqrt{c/(1-\varphi^\star)}}^{+\infty} x^2f(x)\mathrm{d} x -d \Big).
	\end{multline}

	When 
	$\int_{\sqrt{c}}^{+\infty} x^2f(x)\mathrm{d} x \ge d/2$, \cref{thm: optimal_jam_prob} implies that the optimal jamming probability is
	$\varphi^\star=\tilde{\varphi}$ and consequently $\int_{\sqrt{c/(1-\varphi^\star)}}^{+\infty} x^2f(x)\mathrm{d} x -d/2=0$. Therefore, $\mathcal{J}\big((\gamma_{\eta^\star,\varphi^\star}^\star,\eta^\star),\varphi\big)$ is constant for $\varphi \in [0,1)$.
	
	Conversely, if $\int_{\sqrt{c}}^{+\infty} x^2f(x)\mathrm{d} x < d/2$, \cref{thm: optimal_jam_prob} implies that $\varphi^\star=0$. Therefore, $\int_{\sqrt{c/(1-\varphi^\star)}}^{+\infty} x^2f(x)\mathrm{d} x -d/2<0$. In this case, $\varphi=0$ maximizes $\mathcal{J}\big((\gamma_{\eta^\star,\varphi^\star}^\star,\eta^\star),\varphi\big)$.
\end{IEEEproof}

\vspace{5pt}

\Cref{thm:saddle point} summarizes the saddle point strategy for the game between a coordinator jointly designing the transmission and estimation strategy against a proactive jammer.

\vspace{5pt}

\begin{theorem}[saddle point equilibrium for scalar Gaussian sources]\label{thm:saddle point} Given a  Gaussian source $X\sim \mathcal{N}(0,\sigma^2)$, communication and jamming costs $c,d\geq 0$, the saddle point strategy $(\gamma^\star,\eta^\star,\varphi^\star)$ for the remote estimation game with a proactive jammer is given by:	
	\begin{itemize} 
		\item[1)] If $\int_{\sqrt{c}}^{+\infty} x^2f(x)\mathrm{d} x <d/2$, the optimal policies are 
		\begin{align}\label{eq:saddle_pt_11}
			\gamma^\star(x) &= \mathbf{1}(x^2 > c)\\ 
			 \label{eq:saddle_pt_12} 
			\varphi^\star &= 0.
		\end{align}
		
		\item[2)] If $\int_{\sqrt{c}}^{+\infty} x^2f(x)\mathrm{d} x \ge d/2$, the optimal policies are
		\begin{align}\label{eq:saddle_pt_21}
			\gamma^\star(x) &= \mathbf{1}\big((1-\tilde{\varphi})x^2 > c\big) \\ 
			 \label{eq:saddle_pt_22} 
			\varphi^\star &=\tilde{\varphi},
		\end{align}
		where $\tilde{\varphi}$ is the unique solution of \cref{eq: J_d_beta_tilde}.
	\end{itemize}
	In both cases, the optimal estimator is:
	\begin{equation}
	\eta^\star (y) = \begin{cases}
				0, & \text{if} \ \ y \in \{\varnothing,\mathfrak{C}\} \\
				x, & \text{if} \ \ y = x.
			\end{cases}
		\end{equation}
\end{theorem}

\vspace{5pt}

\begin{IEEEproof} We need to consider two cases.
	
\textbf{Case 1}  -- Assume that $2\int_{\sqrt{c}}^{+\infty} x^2f(x)\mathrm{d} x <d$. If the jammer chooses not to block the channel, i.e., $\varphi^\star=0$, \cref{prop:threshold_n_1} implies that the corresponding optimal transmission strategy is
	$
	\gamma^\star(x) = \mathbf{1}(x^2 > c). 
	$
	Under this pair of jamming and transmission policies, \cref{thm: optimal_estimator_n_1} yields that $\hat{x}^\star_0=0$ and $\hat{x}_1^\star=0$. Therefore,
	\begin{equation}
	\mathcal{J}\big( (\gamma^\star,\eta^\star),\varphi^\star \big) \le \mathcal{J}\big( (\gamma,\eta),\varphi^\star \big).
	\end{equation}
	If the optimal transmission strategy is $\gamma^\star$ and the optimal estimator is $\eta^\star$,  \cref{col:LHS_saddle} implies that
	\begin{equation}
	\mathcal{J}\big( (\gamma^\star,\eta^\star),\varphi \big) \le \mathcal{J}\big( (\gamma^\star,\eta^\star),\varphi^\star \big).
	\end{equation}
	
	\textbf{Case 2} -- Assume that $2\int_{\sqrt{c}}^{+\infty} x^2f(x)\mathrm{d} x \ge d$. If the jammer blocks the channel with probability $\tilde{\varphi}$, \cref{prop:threshold_n_1} implies that the corresponding optimal transmission strategy is
	$
	\gamma^\star(x) = \mathbf{1}\big((1-\tilde{\varphi})x^2 > c\big).
	$
	Under this pair of jamming and transmission policies, \cref{thm: optimal_estimator_n_1} yields that $\hat{x}^\star_0=0$ and $\hat{x}_1^\star=0$. Therefore, 
	\begin{equation}
	\mathcal{J}\big( (\gamma^\star,\eta^\star),\varphi^\star \big) \le \mathcal{J}\big( (\gamma,\eta),\varphi^\star \big).
	\end{equation}
	If the optimal transmission strategy is $\gamma^\star$ and the optimal estimator is $\eta^\star$,  \cref{col:LHS_saddle} implies that
	\begin{equation}
	\mathcal{J}\big( (\gamma^\star,\eta^\star),\varphi \big) \le \mathcal{J}\big( (\gamma^\star,\eta^\star),\varphi^\star \big).
	\end{equation}
\end{IEEEproof}
	
}

	\begin{remark}
		Notice that in case 2 of \cref{thm:saddle point}, the ratio $c/(1-\tilde{\varphi})$ is constant for any given value of $d>0$, which is determined by solving \cref{eq: J_d_beta_tilde}. Therefore, the optimal transmission policy is also uniquely determined by $d$.
	\end{remark}

\subsection{Reactive jamming of point-to-point collision channels} \label{sec:ReactiveJammer}

\begin{figure}[t!]
	\centering
	\includegraphics[width=0.75\columnwidth]{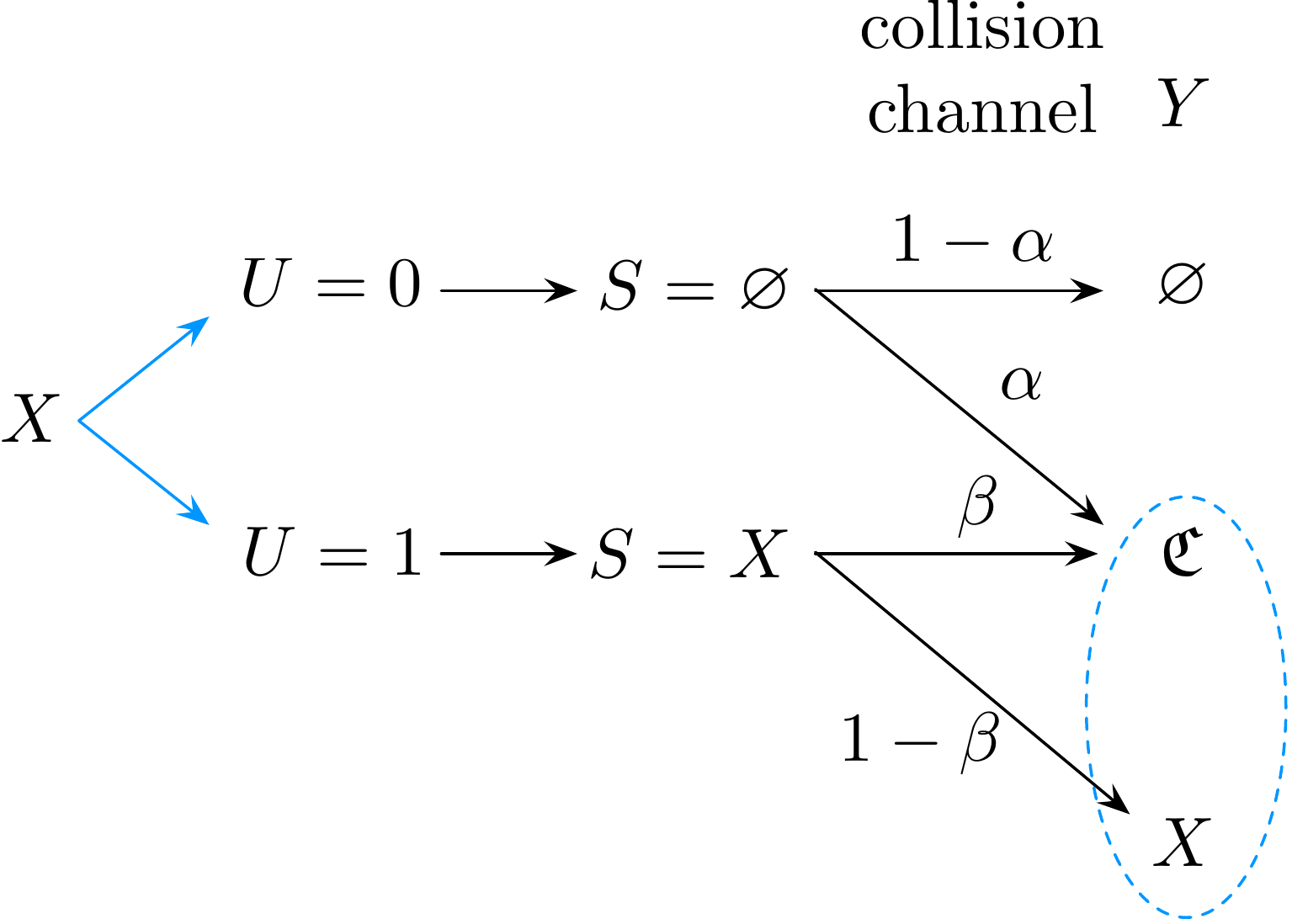}
	\caption{Signaling channel between the sensor and the receiver. The jammer controls the transition probabilities $\alpha$ and $\beta$. When $\alpha=\beta=0$, the channel is noiseless, i.e., the receiver can unequivocally decode whether $U=1$ or $U=0$ from the ouput signal $Y$.}
	\label{fig:channel}
\end{figure}

In this section, we consider the case in which the attacker can sense whether the channel is occupied or not. Notice that we allow the reactive jammer to block the channel even when the sensor is not transmitting. To the best of our knowledge, the existing literature on reactive jamming attacks precludes that possibility. However, there is a reason why the jammer may engage in such counter-intuitive behavior: when the jammer only blocks a transmitted signal, it creates a noiseless binary signaling channel between the transmitter and the receiver, which may be exploited by the coordinator. If the jammer is allowed to ``block'' the channel when the user is not transmitting, such binary signaling channel will no longer be noiseless because there will be uncertainty if the decision variable at the transmitter is zero or one. This scenario is illustrated in \cref{fig:channel}.

\vspace{5pt}


\begin{proposition}
	\label{prop: best_threshold_reactive}
	For a fixed jamming policy parametrized by $\varphi\in [0,1]^2$, and a fixed estimation policy $\eta$ parametrized by $\hat{x}\in \mathbb{R}^{2m}$, the optimal transmission policy is:
	\begin{multline}
		\gamma_{\eta,\varphi}^\star(x) = \mathbf{1}\big( \beta \|x-\hat{x}_1\|^2  + c - d \beta < \\ \alpha \|x-\hat{x}_1\|^2  + (1-\alpha) \|x-\hat{x}_0\|^2 - d \alpha  \big).
	\end{multline}
\end{proposition}

\vspace{5pt}

\begin{IEEEproof} 
	For a reactive jammer, the random variables $X$ and $J$ are conditionally independent given $U$. Using the law of total expectation, and employing the estimation policy in \cref{eq:estimator1}, the cost function can be expressed as
	\begin{multline} \label{eq:cost_reactive}
		\mathcal{J}\big( (\gamma,\eta),\varphi \big) = 
		\int_{\mathbb{R}^{m}} \big[\beta \|x-\hat{x}_1\|^2  + c - d \beta \big]\gamma(x)f(x)\mathrm{d} x +\\
		\int_{\mathbb{R}^{m}} \big[\alpha \|x-\hat{x}_1\|^2  + (1-\alpha) \|x-\hat{x}_0\|^2 - d \alpha\big]\big((1-\gamma(x)\big)f(x)\mathrm{d} x.
	\end{multline}
	
	For fixed $\hat{x} \in\mathbb{R}^{2m}$ and $\varphi\in[0,1]^2$, the transmission policy $\gamma$ that minimizes \cref{eq:cost_reactive} is obtained by comparing the arguments of the two integrals as follows:  $x \in \{\xi \mid \gamma^\star_{\eta,\varphi}(\xi) = 1\}$ if and only if 
	\begin{equation}
		\beta \|x-\hat{x}_1\|^2  + c - d \beta < \\ \alpha \|x-\hat{x}_1\|^2  + (1-\alpha) \|x-\hat{x}_0\|^2 - d \alpha.
	\end{equation}
\end{IEEEproof}

\vspace{5pt}

Given the optimal transmitter's strategy in \cref{prop: best_threshold_reactive}, the objective function becomes
\begin{multline}\label{eq:cost_reactive_new}{}
	\mathcal{J}\big((\gamma_{\eta,\varphi}^{\star},\eta),\varphi\big) = \mathbf{E}\Big[ \min\big\{\beta\|X-\hat{x}_1\|^2 +c-d\beta , \\ \alpha\|X-\hat{x}_1\|^2 + (1-\alpha)\|X-\hat{x}_0\|^2 -d\alpha \big\} \Big] \Equaldef \tilde{\mathcal{J}}(\hat{x},\varphi).
\end{multline}

Therefore, the coordinator wants to minimize $\tilde{\mathcal{J}}(\hat{x},\varphi)$ over $\hat{x}\in\mathbb{R}^{2m}$ and the jammer wants to maximize it over $\varphi\in [0,1]^2$. As in \cref{sec:pt2pt}, for fixed $\hat{x}\in\mathbb{R}^{2m}$, $\tilde{\mathcal{J}}$ is a concave function of $\varphi$ for any pdf $f$. However, for fixed $\varphi\in [0,1]^2$, $\tilde{\mathcal{J}}$ is non-convex in $\hat{x}$. Unfortunately, the structure of \cref{eq:cost_reactive_new} does not allow us to use the same techniques to find a saddle point equilibrium for the proactive jammer. It is also not clear if saddle point solutions even exist.

From the remainder of this section, we assume that the coordinator and the jammer are solving the following minimax optimization problem\footnote{A solution for the minimax problem corresponds to finding a \textit{security} (or robust) policy for the coordinator \cite{Hespanha:2017}.}:
\begin{equation}\label{eq:game}
	\min_{\hat{x}\in \mathbb{R}^{2m}} \max_{\varphi \in [0,1]^2} \tilde{\mathcal{J}}(\hat{x},\varphi),
{}\end{equation}
where $\tilde{\mathcal{J}}(\hat{x},\varphi)$ is given by \cref{eq:cost_reactive_new}.

\vspace{5pt}

A useful alternative to the saddle point equilibrium are the solutions that satisfy the first-order stationarity conditions of the minimization and the maximization problems, yielding in a larger class of policies, called \textit{first order Nash equilibria} (FNE) \cite{Ostrovskii:2021,Nouiehed:2019,Facchinei:2003,Lin:2020}. 


\vspace{5pt}

\begin{definition}[Approximate First order Nash equilibrium]
	Let $\varepsilon >0$. A pair of policies $(\hat{x}^{\star},\varphi^{\star}) \in \mathbb{R}^{2m}\times [0,1]^2$ is an approximate First-order Nash-equilibrium ($\varepsilon$-FNE) of the game if 
	\begin{equation} \label{eq:FNEcond}
		\|\nabla_{\hat{x}}\tilde{\mathcal{J}}(\hat{x}^{\star},\varphi^{\star})\|_2 \le \varepsilon
	\end{equation}
	and
	\begin{equation}\label{eq:LP_condition}
		\max_{\varphi \in[0,1]^2}\langle \nabla_{\varphi}\tilde{\mathcal{J}}(\hat{x}^{\star},\varphi^{\star}),\varphi-\varphi^\star\rangle \leq \varepsilon.
	\end{equation}
\end{definition}

\vspace{5pt}

%

\begin{proposition}\label{prop:gradients}
	The function $\tilde{\mathcal{J}}(\hat{x},\varphi)$ admits the following subgradients with respect to $\hat{x}$ and $\varphi$: 
	\begin{multline} \label{eq:gradient_x_hat}
		\nabla_{\hat{x}}\tilde{\mathcal{J}}(\hat{x},\varphi) 
		= \textbf{E}
		\Bigg[ 
		\begin{bmatrix}
			0_{m\times 1} \\
			-2 \beta (X-\hat{x}_1)
		\end{bmatrix}\mathbf{1}(\gamma_{\eta,\varphi}^\star(X)=1)  \\
		+
		\begin{bmatrix}
			-2(1-\alpha) (X-\hat{x}_0)  \\
			-2 \alpha (X-\hat{x}_1) 
		\end{bmatrix} \mathbf{1}(\gamma_{\eta,\varphi}^\star(X)=0)
		\Bigg]
	\end{multline}
	and
	\begin{multline} \label{eq:gradient_theta}
		\nabla_{\varphi}\tilde{\mathcal{J}}(\hat{x},\varphi) 
	=\textbf{E}\Bigg[ 
		 \begin{bmatrix} 
			0 \\
			\|X-\hat{x}_1\|^2 - d 
		\end{bmatrix}\cdot\mathbf{1}(\gamma_{\eta,\varphi}^\star(X)=1) \\
		+
		\begin{bmatrix}
			\|X-\hat{x}_1\|^2  - \|X-\hat{x}_0\|^2 - d   \\
			0
		\end{bmatrix} \mathbf{1}(\gamma_{\eta,\varphi}^\star(X)=0)\Bigg].
	\end{multline}
\end{proposition}

\vspace{5pt}

\begin{IEEEproof}
	This result follows from the Leibniz rule. 
\end{IEEEproof}
\vspace{5pt}

Problems in the form of \cref{eq:game} where the inner optimization problem is concave and the outer optimization problem is non-convex have been studied under assumptions on the gradients being Lipschitz continuous \cite{Nouiehed:2019,Lin:2020}. Under such conditions an algorithm known as the (Projected) Gradient Ascent-Descent (GAD) converges to an $\varepsilon$-FNE. However, the gradients in \cref{eq:gradient_x_hat,eq:gradient_theta} 
are not Lipschitz continuous. We will resort to an alternative algorithm that leverages the structure of a difference of convex decomposition present in our problem.

\vspace{5pt}

	\subsubsection{Optimization algorithm for a reactive jammer}
	
	To obtain a pair of $\varepsilon$-FNE to the problem in \cref{eq:game}, we alternate between a \textit{projected gradient ascent} (PGA) step for the inner optimization problem; and a \textit{convex-concave procedure} (CCP) step for the outer optimization problem.

	We start with the description of the PGA step at a point $(\hat{x}^{(k)},\varphi^{(k)})$:
	\begin{equation}
		\varphi^{(k+1)}= \mathcal{P}_{[0,1]^2}\big(\varphi^{(k)} + \lambda_k\, \nabla_{\varphi}\tilde{\mathcal{J}}(\hat{x}^{(k)},\varphi^{(k)})\big),
	\end{equation}
	where $\{\lambda_k\}$ is a step-size sequence (e.g. $\lambda_k=0.1/\sqrt{k}$) and the projection operator is defined as  
	\begin{equation}
	\mathcal{P}_{[0,1]^2}(\varphi)\Equaldef \min_{\bar{\varphi} \in [0,1]^2} \| \bar{\varphi}-\varphi \|_2,
\end{equation}
	 which is equal to
	\begin{equation}
		\mathcal{P}_{[0,1]^2} \bigg( 
		\begin{bmatrix}
			\alpha \\
			\beta
		\end{bmatrix} \bigg) = \bigg[ 
		\begin{array}{c}
			\max\big\{0, \min\{1, \alpha\} \big\} \\
			\max\big\{0, \min\{1, \beta\} \big\}
		\end{array} \bigg].
	\end{equation}
	

	To update $\hat{x}^{(k)}$ for a fixed $\varphi^{(k+1)}$, we use the property that \cref{eq:cost_reactive_new} can be decomposed as a difference of convex functions (DC decomposition). Using the  
	DC decomposition we obtain a specialized descent algorithm, which is guaranteed to converge to stationary points of \cref{eq:cost_reactive_new} for a fixed $\varphi^{(k+1)}$.
	Because the CCP uses more information about the structure of the objective function than standard Gradient Descent methods, it often leads to faster convergence \cite{Yuille:2003,Lipp:2016}.   

\begin{algorithm}[t]
	\caption{PGA-CCP algorithm}
	\label{alg: I}
	\begin{algorithmic}[1]
		\REQUIRE  PDF $f$, transmission cost $c$, jamming cost $d$   
		\ENSURE   Estimated result $\hat{x}^{\star}$ and $\varphi^{\star}$
		\STATE
		Initialize 
		$k\gets0,$ $\varepsilon,$ $\hat{x}^{(0)}$ and $\varphi^{(0)}$
		\REPEAT
		\STATE $\varphi^{(k+1)}=   \mathcal{P}_{[0,1]^2}\big(\varphi^{(k)} + \lambda_k\,\nabla_{\varphi}\tilde{\mathcal{J}}(\hat{x}^{(k)},\varphi^{(k)})\big)$
		\STATE $\hat{x}^{(k+1)} = \mathcal{A}^\dagger(\varphi^{(k+1)}) \, g(\hat{x}^{(k)},\varphi^{(k+1)}) +\mu$
		\STATE $k \gets k+1$
		\UNTIL $\varepsilon$-FNE conditions (\cref{eq:FNEcond,eq:LP_condition}) are satisfied 
	\end{algorithmic}
\end{algorithm}

Notice that:
\begin{equation}
	\tilde{\mathcal{J}}(\hat{x},\varphi)= \mathcal{F}(\hat{x},\varphi)-\mathcal{G}(\hat{x},\varphi),
\end{equation}
where
\begin{multline}
	\mathcal{F}(\hat{x},\varphi) \Equaldef (1-\alpha) \|\hat{x}_0\|^2 +(\alpha+\beta) \|\hat{x}_1\|^2 \\+(1+\beta)\big(\trace(\Sigma)+\|\mu\|^2\big)+c-d(\alpha+\beta)\\
	-2\big[(\beta+\alpha)\hat{x}_1 + (1-\alpha)\hat{x}_0\big]^\mathsf{T}\mu,
\end{multline}
and
\begin{multline}
	\mathcal{G}(\hat{x},\varphi) \Equaldef \mathbf{E}\Big[ \max\big\{\beta\|X-\hat{x}_1\|^2 +c-d\beta , \\ \alpha\|X-\hat{x}_1\|^2 + (1-\alpha)\|X-\hat{x}_0\|^2 -d\alpha \big\} \Big].
\end{multline}

The CCP for computing a local minima for the outer optimization problem is given by
\begin{equation} \label{eq: x_optimization_problem}
	\hat{x}^{(k+1)}= \arg \min_{\hat{x}} \left\{ \mathcal{F}(\hat{x},\varphi^{(k+1)})- 	\mathcal{G}_{\mathrm{affine}}(\hat{x},\varphi^{(k+1)};\hat{x}^{(k)}) \right\},
\end{equation}
where $\mathcal{G}_{\mathrm{affine}}(\hat{x},\varphi^{(k+1)};\hat{x}^{(k)})$ is the affine approximation of $\mathcal{G}(\hat{x},\varphi^{(k+1)})$ with respect to $\hat{x}$ at $\hat{x}^{(k)}$, while keeping $\varphi^{(k+1)}$ fixed, i.e.,
\begin{multline}
	\mathcal{G}_{\mathrm{affine}}(\hat{x},\varphi^{(k+1)};\hat{x}^{(k)})= \mathcal{G}(\hat{x}^{(k)},\varphi^{(k+1)})  \\+g(\hat{x}^{(k)},\varphi^{(k+1)})^{\mathsf{T}} (\hat{x}-\hat{x}^{(k)})
\end{multline}
and $g(\hat{x},\varphi)$ is the gradient of $\mathcal{G}(\hat{x},\varphi)$  with respect to $\hat{x}$.

Because $\mathcal{F}$ is a quadratic function of $\hat{x}$ for a fixed $\varphi$, we may use the first-order necessary optimality condition of problem \cref{eq: x_optimization_problem} to find the recursion for $\hat{x}^{(k+1)}$ in closed form:
\begin{equation}
	\nabla_{\hat{x}} \mathcal{F}(\hat{x}^{(k+1)},\varphi^{(k+1)})=g(\hat{x}^{(k)},\varphi^{(k+1)}).
\end{equation}
The partial gradient of $\mathcal{F}(\hat{x},\varphi)$ with respect to $\hat{x}$ is
\begin{equation}
	\nabla_{\hat{x}} \mathcal{F}(\hat{x},\varphi)=\left[ 
	\begin{array}{c}
		2(1-\alpha) (\hat{x}_0-\mu) \\
		2(\alpha+\beta) (\hat{x}_1-\mu)
	\end{array} \right].
\end{equation}

The partial gradient of $\mathcal{G}(\hat{x},\varphi)$  with respect to $\hat{x}$ is
\begin{multline}
	g(\hat{x},\varphi) 
	= \textbf{E}
	\Bigg[ 
	\begin{bmatrix}
		0_{m\times 1} \\
		-2 \beta (X-\hat{x}_1)
	\end{bmatrix}\mathbf{1}(\gamma_{\eta,\varphi}^\star(X)=0)  \\
	+
	\begin{bmatrix}
		-2(1-\alpha) (X-\hat{x}_0)  \\
		-2 \alpha (X-\hat{x}_1) 
	\end{bmatrix} \mathbf{1}(\gamma_{\eta,\varphi}^\star(X)=1)
	\Bigg].
\end{multline}

Finally, define $\mathcal{A}:[0,1]^2\rightarrow \mathbb{R}^{2 m \times 2 m}$ as
\begin{equation}
	\mathcal{A}(\varphi) \Equaldef \left[ 	\begin{array}{cc}
		2(1-\alpha) I_{m\times m} & 0_{m\times m}\\
		0_{m\times m} & 2(\alpha+\beta) I_{m\times m}
	\end{array} \right],
\end{equation}
and $\mathcal{A}^\dagger$  denotes its Moore-Penrose pseudo-inverse. Then, the update of CCP can be compactly represented as
\begin{equation}
	\hat{x}^{(k+1)} = \mathcal{A}^\dagger\big{(}\varphi^{(k+1)}\big{)} \, g\big{(}\hat{x}^{(k)},\varphi^{(k+1)}\big{)} +\mu.
\end{equation}


\begin{figure}[t]
	\centering
	\includegraphics[width=0.9 \columnwidth]{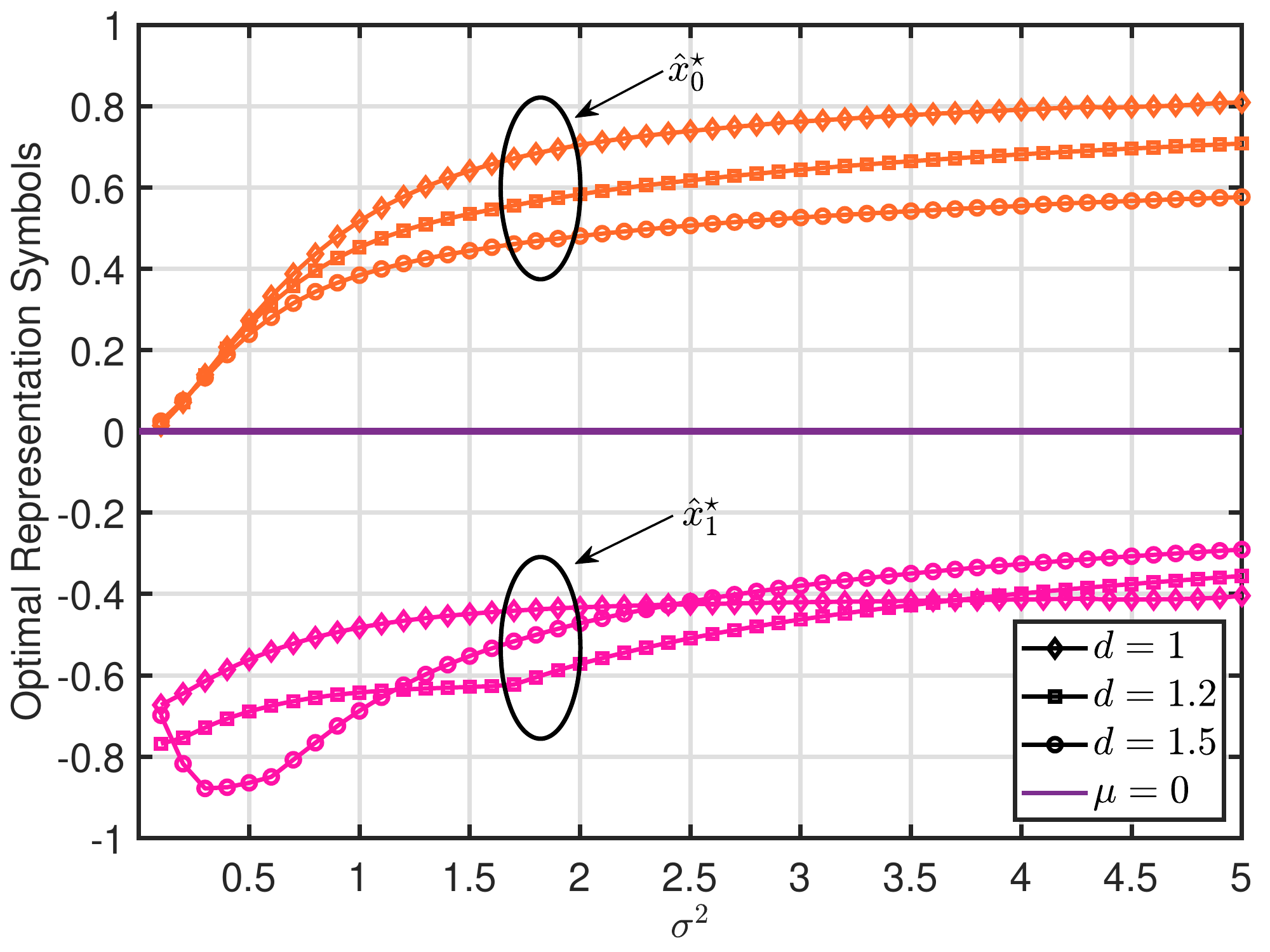}
	\caption{Optimal representation symbols $\hat{x}_0^\star$ and $\hat{x}_1^\star$ as a function of $\sigma^2$ for different jamming cost $d$, where $X \sim \mathcal{N}(0,\sigma^2)$ and $c=1$. }
	\label{fig:Optimal_xhat_vs_Var_all}
\end{figure}

\begin{figure}[t]
	\centering
	\includegraphics[width=0.9 \columnwidth]{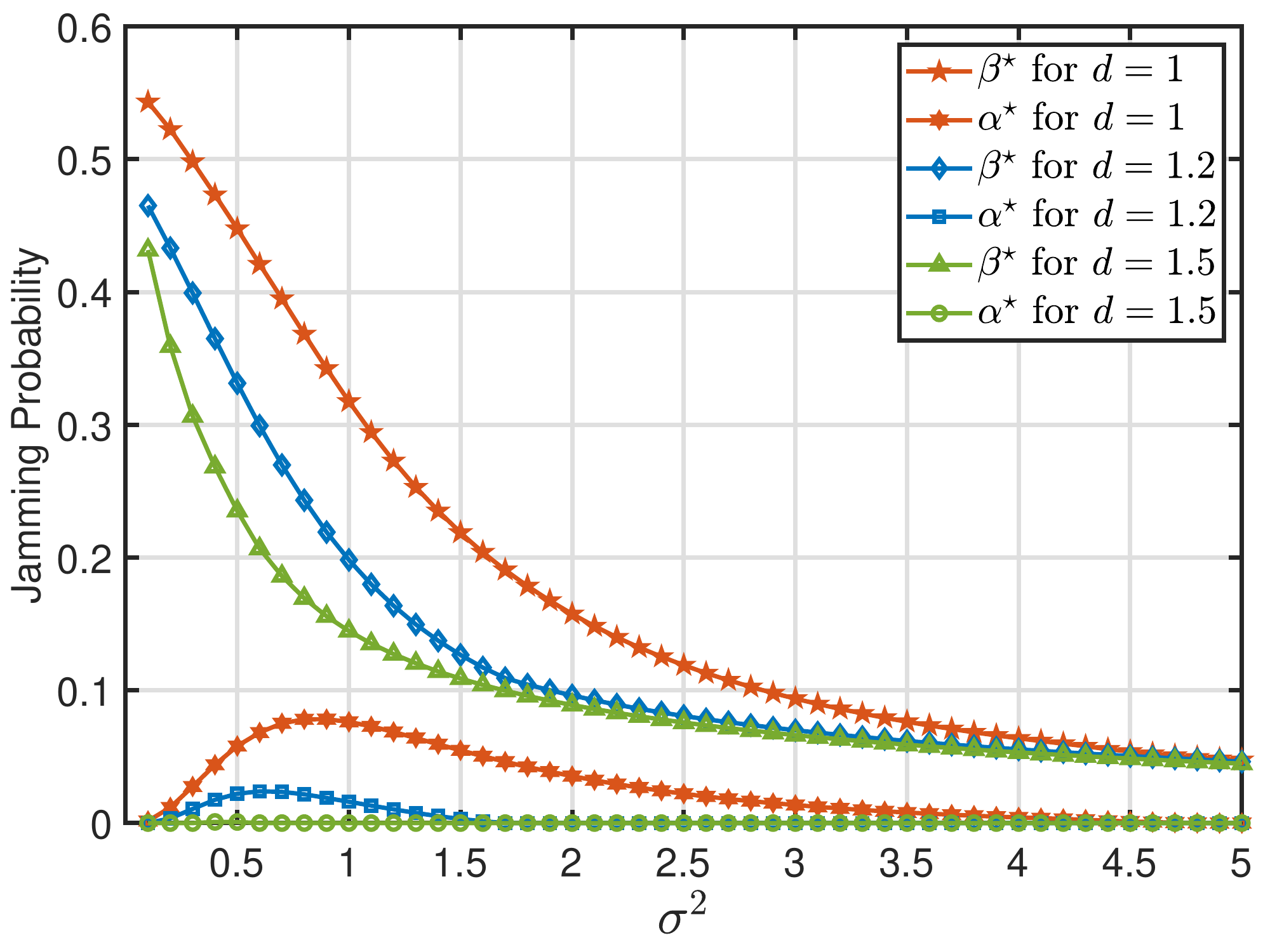}
	\caption{Optimal jamming probabilities $\alpha^\star$ and $\beta^\star$ as a function of $\sigma^2$ for different jamming cost $d$, where $X \sim \mathcal{N}(0,\sigma^2)$ and $c=1$. }
	\label{fig:Optimal_alphabeta_vs_Var_all}
\end{figure}

%
%


\subsection{Numerical results}
 In this subsection, we provide policies that satisfy $\varepsilon$-FNE. Convergence is studied using the ``performance index'' defined below  
\begin{multline}
	{\textup{FNE}}(\hat{x}^{(k)},\varphi^{(k)})  \Equaldef \max \big\{\|\nabla_{\hat{x}}\tilde{\mathcal{J}}(\hat{x}^{(k)},\varphi^{(k)})\|_2,\\  \max_{\varphi \in[0,1]^2}\langle \nabla_{\varphi}\tilde{\mathcal{J}}(\hat{x}^{(k)},\varphi^{(k)}),\varphi-\varphi^{(k)}\rangle \big\}.
\end{multline}
In this section, ``optimality'' is in the $\varepsilon$-FNE sense.
We begin by presenting the optimal estimation policies for one-dimensional observations. \cref{fig:Optimal_xhat_vs_Var_all} shows the optimal representation symbols $\hat{x}_0^\star$ and $\hat{x}_1^\star$ as a function of $\sigma^2$ for different jamming cost $d$, where $X \sim \mathcal{N}(0,\sigma^2)$, $c=1$ and $\varepsilon=10^{-5}$.  Notice that the representation symbols obtained for the collision and no-transmission in the presence of the reactive jammer are always distinct and neither is equal to the mean $0$. This is in contrast with the the proactive jammer case, in which $\hat{x}^\star_0=\hat{x}_1^\star= 0$. Therefore, the assumption of a fixed receiver with $\hat{x}^\star_0=\hat{x}_1^\star= 0$, as in \cite{gupta2012dynamic,gupta2016
}, leads to a loss of optimality.  



We then present the optimal jamming policies for one-dimensional observations. \Cref{fig:Optimal_alphabeta_vs_Var_all}
shows the optimal jamming probabilities $\alpha^\star$ and $\beta^\star$ as a function of $\sigma^2$ for different jamming cost $d$ with $X \sim \mathcal{N}(0,\sigma^2)$, $c=1$ and $\varepsilon=10^{-5}$. Notice that the optimal jamming probabilities decrease as $d$ increases. Besides, the optimal jamming probability when the sensor does not transmit can be nonzero when $d=1$ and $d=1.2$, where the jammer aims to deceive the estimator into thinking there has been a transmission that has been blocked. However, when $d=1.5$ the optimal jamming probability when the sensor does not transmit is zero for all $\sigma^2$ since the jamming cost is high and it is not worth it to  deceive the estimator. 

\begin{figure}[htb]
	\centering
	\subfloat{
		\includegraphics[width=0.9 \columnwidth]{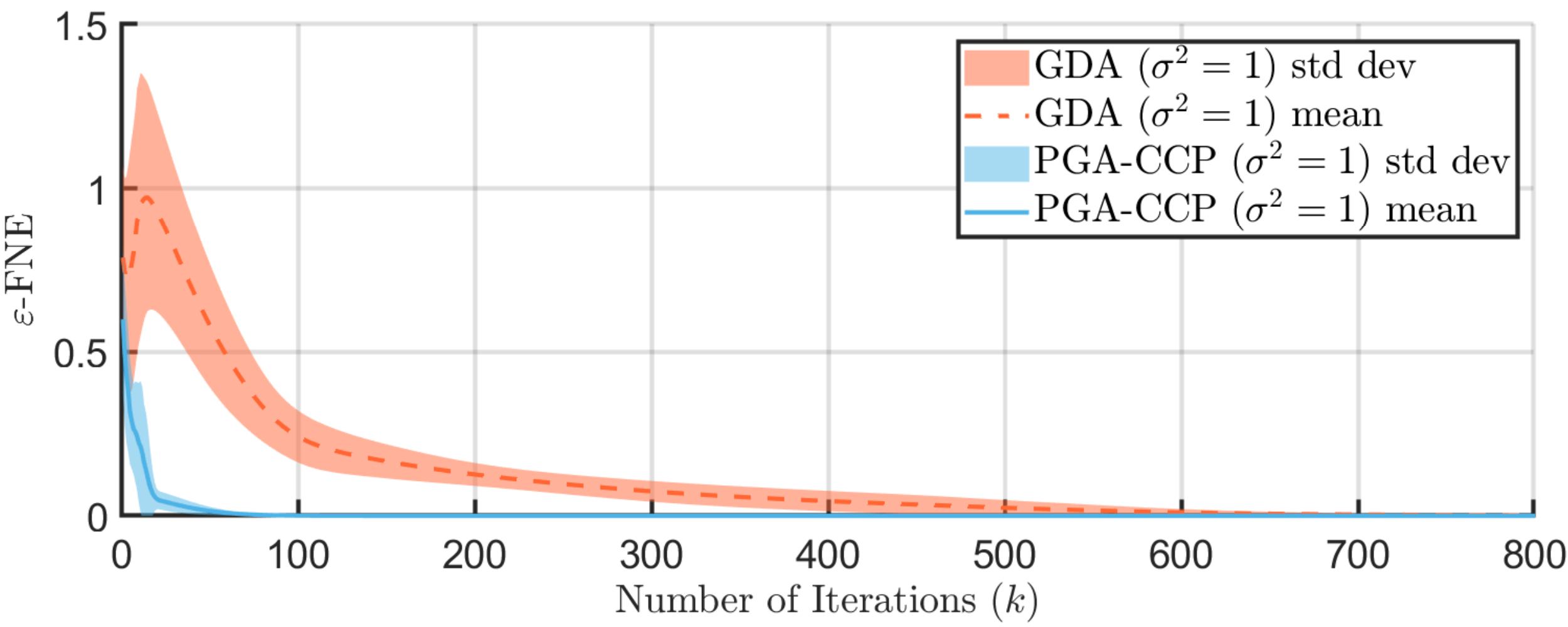}}
	\hfil
	\subfloat{
		\includegraphics[width=0.9 \columnwidth]{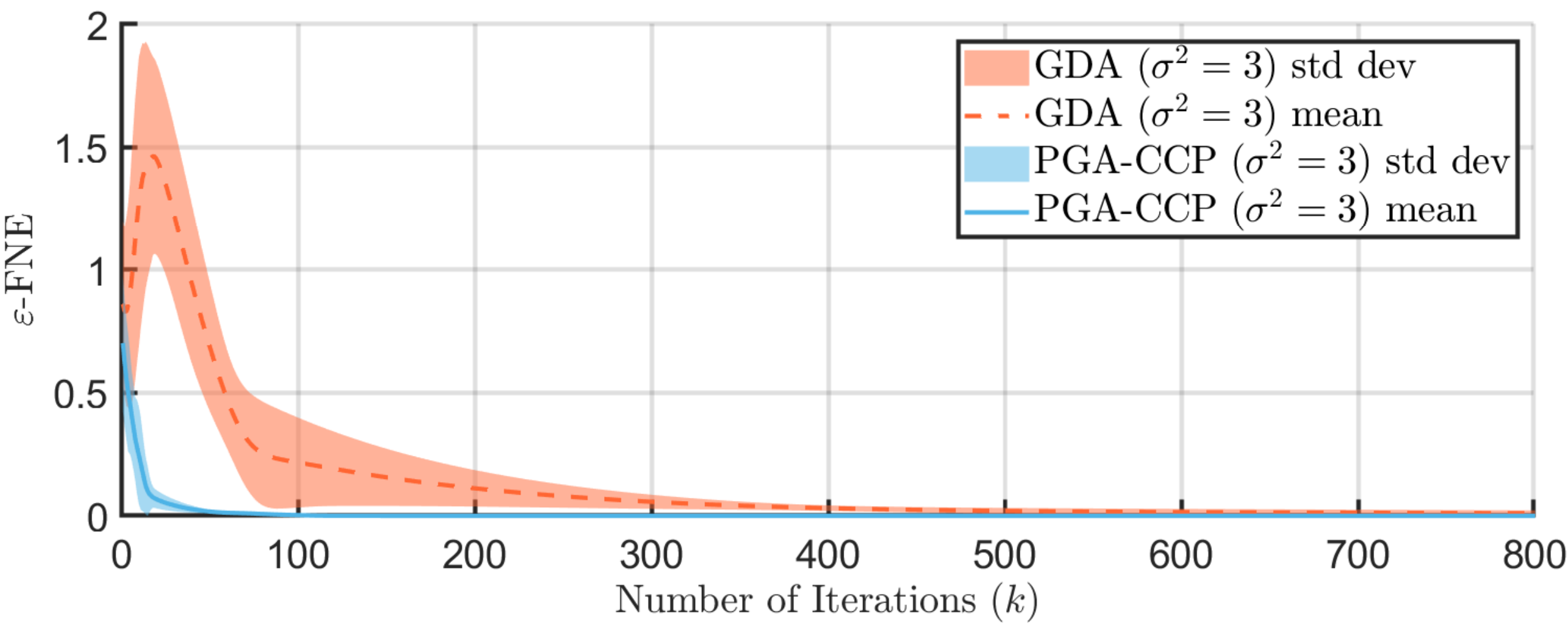}}
	\hfil
	\subfloat{
		\includegraphics[width=0.9 \columnwidth]{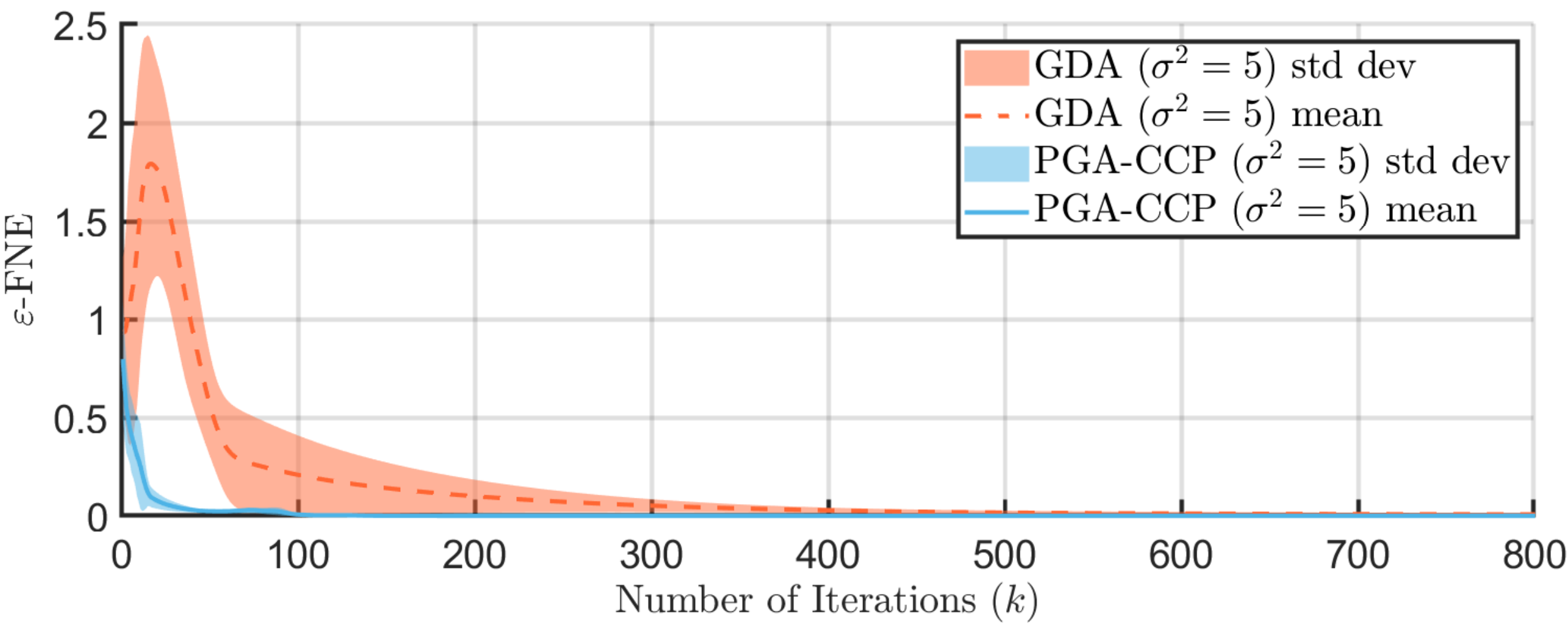}}
	\caption{Convergence curves of PGA-CCP vs. GDA for different variance $\sigma^2$, where $m=1$, $c=1, d=1$, and $X \sim \mathcal{N}(0,\sigma^2)$. \textbf{Top:} $\sigma^2=1$; \textbf{Middle:} $\sigma^2=3$; \textbf{Bottom:} $\sigma^2=5$. The results are obtained by taking the average of 100 Monte Carlo simulations.}
\label{fig:comparison}
\end{figure}

\begin{figure}[t!]
\centering
\subfloat{
	\includegraphics[width=0.9 \columnwidth]{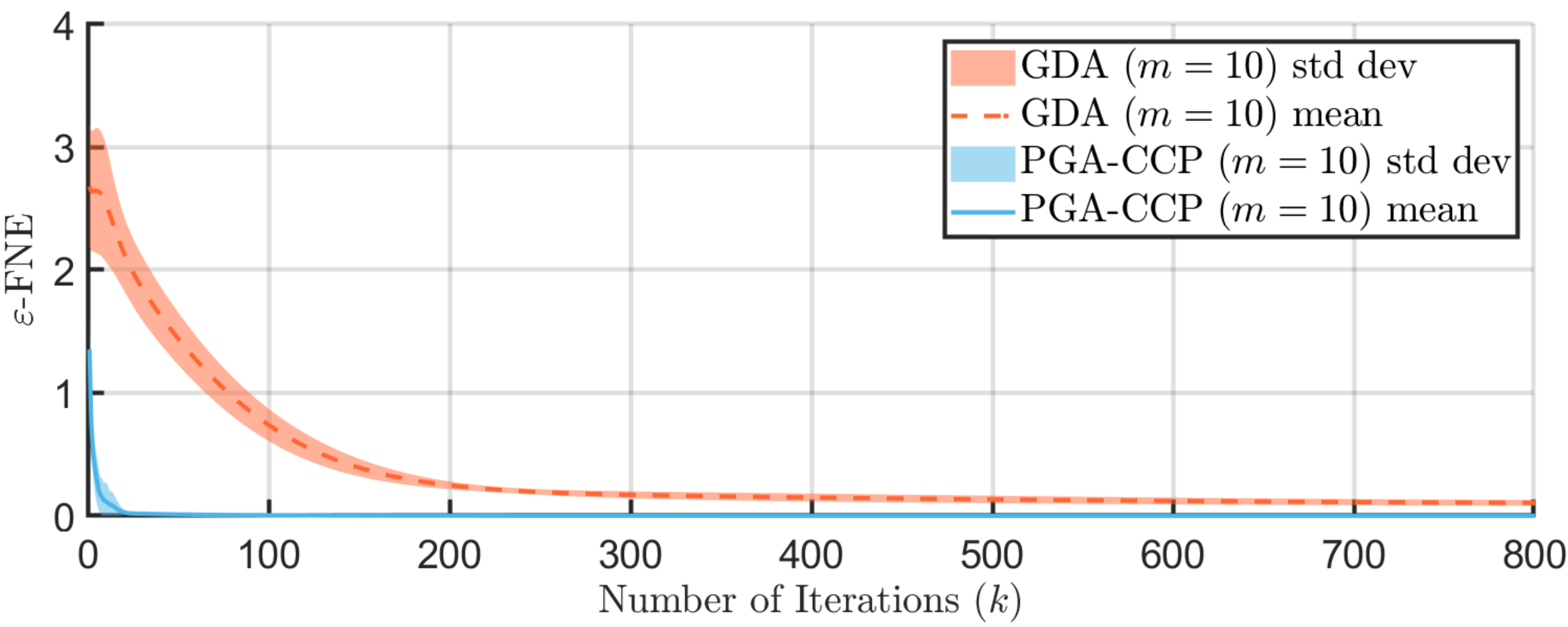}}
\hfil
\subfloat{
	\includegraphics[width=0.9 \columnwidth]{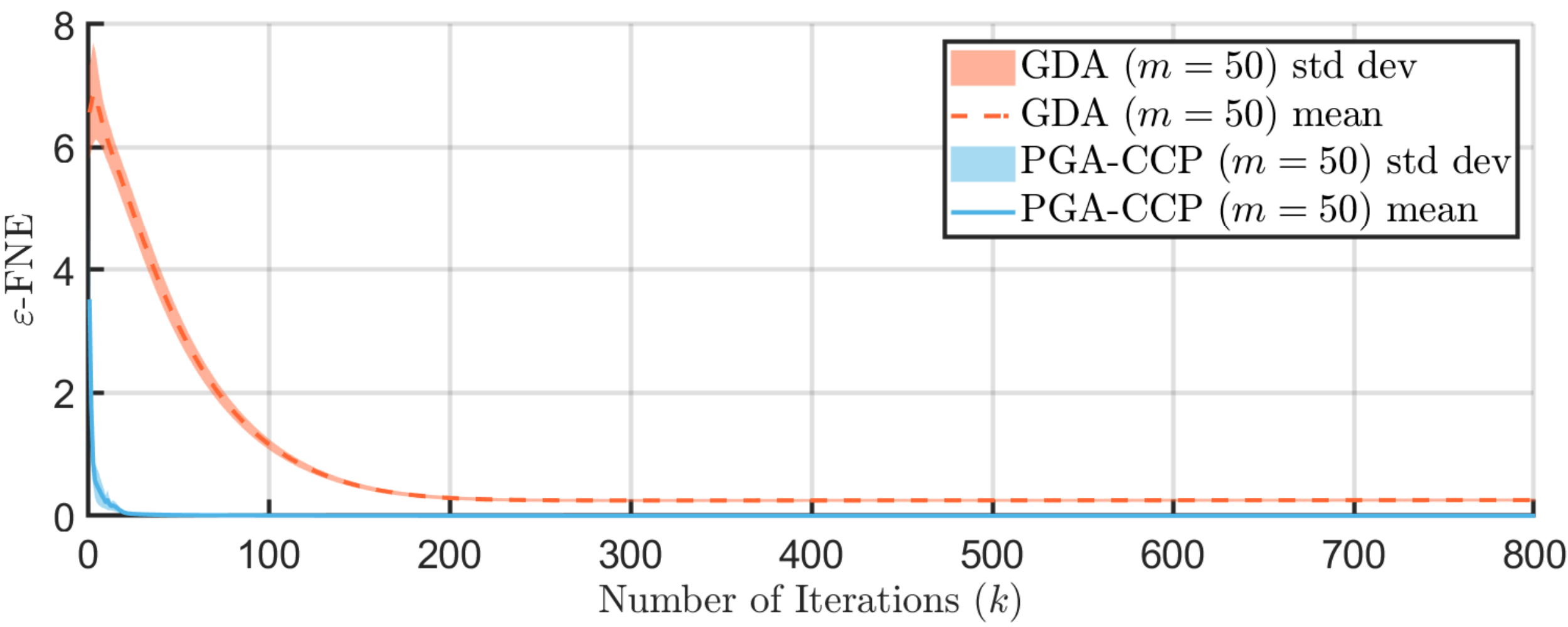}}
\hfil
\subfloat{
	\includegraphics[width=0.9 \columnwidth]{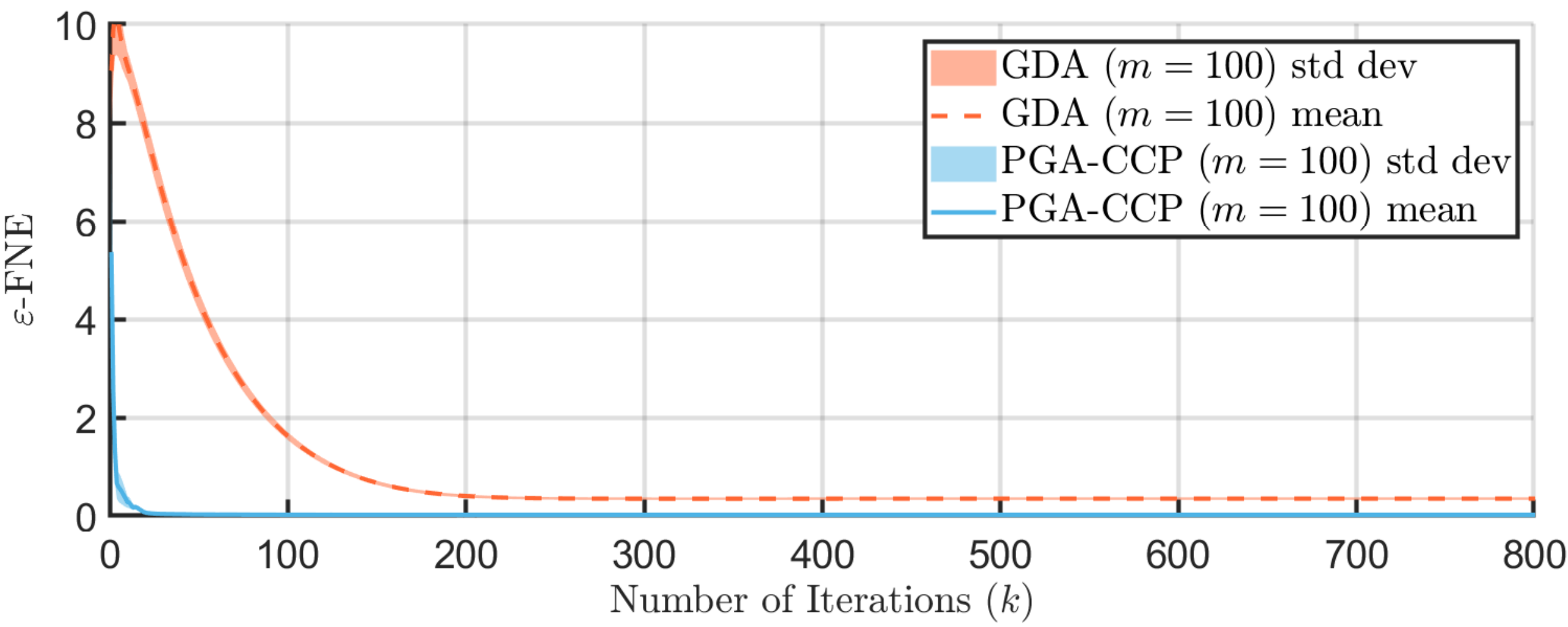}}
\caption{Convergence curves of PGA-CCP vs. GDA for multidimensional state, where $c=1,d=1$, and $X \sim \mathcal{N}(0_{m},I_{m\times m})$.  \textbf{Top:}  $m=10$; \textbf{Middle:} $m=50$; \textbf{Bottom:} $m=100$. The sample size for the estimation of gradients is $10^4$.}
\label{fig:comparison2}
\end{figure}

We also compare the performance of our proposed PGA-CCP and the traditional GDA algorithms. \Cref{fig:comparison} presents the convergence curves of PGA-CCP vs. GDA for different values of the variance $\sigma^2$, where $c=1,d=1$, and $X \sim \mathcal{N}(0,\sigma^2)$. In this study, we set the step size of PGA-CCP as $\lambda=0.1$ and the step sizes for GA and GD in GDA as $\lambda_{\rm{GA}}=0.1$ and $\lambda_{\rm{GD}}=0.01$, respectively. These are values consistent with the ones suggested by the analysis in \cite{Lin:2020}. We performed 100 Monte Carlo simulations for each algorithm with random initial conditions. The results indicate that PGA-CCP converges more than six times faster than GDA. Furthermore, GDA oscillates more with the increase of $\sigma^2$ while PGA-CCP decreases steadily with a small standard deviation from the mean of the sample paths.  

We proceed by presenting the simulation results for multi-dimensional observations. \Cref{fig:comparison2} shows the convergence to $\varepsilon$-FNE for multidimensional observations, with $c=1$ and $d=1$. We make 100 Monte Carlo simulations for each algorithm. For each Monte Carlo simulation, the expectations used in \cref{alg: I} are approximated by the average of $10^4$ samples drawn from $X \sim \mathcal{N}(0_{m},I_{m\times m})$. Notice that PGA-CCP converges quickly to zero while GDA does not even converge to a $0.1$-FNE when $m=10$ and $m=50$. When the dimension of measurements is $m=100$, PGA-CCP can achieve $0.02$-FNE while GDA does not even converge to $0.3$-FNE. Therefore, the numerical examples herein show that our heuristic algorithm is promising relative to GDA especially for high-dimensional remote estimation problems\footnote{The code used to obtain all the examples in this paper is available at GitHub \url{https://github.com/mullervasconcelos/IEEE-TAC2023.git}.}.

\section{Large-scale networks}
In this section, we consider the remote estimation problem over large-scale networks in the presence of a proactive jammer, where the network consists of countably infinitely many legitimate transmitters that can support a fraction $\bar{\kappa} \in (0,1)$ of packets per time-slot transmitted simultaneously, i.e., $\lim_{n \to \infty}  \kappa(n)/n=\bar{\kappa}$. To keep the notation simple, we will only consider the scalar observation case. However, it is straightforward to extend our results to the vector case using the techniques developed in Appendix \ref{appendix: p2p_vector}. Our goal is to obtain an expression for the limiting objective function when $n$ approaches infinity and characterize its saddle point equilibrium. Before that, we will provide the objective function of the remote estimation problem over medium-scale networks, where $n$ is finite.

\subsection{Objective function in medium-scale networks}

In this section, we consider the remote estimation problem over the medium-scale networks, which consists of $n<\infty$ transmitters and support $\kappa(n)<n$ simultaneous packets. Let $\{U_i\}_{i=1}^n$ be the collection of transmission decisions at the sensors. For a given realization of $\{U_i\}_{i=1}^n \in \{0,1\}^n$, define $\mathbb{T}=\{i \mid U_i=1\}$ as the index set of all transmitting sensors. Given the channel input $\{S_i\}_{i=1}^n$ and the jammer's decision $J$, the output of the collision channel of capacity $\kappa(n)<n$ is given by

\begin{equation}
	Y = \begin{cases}
		\varnothing, & \text{if} \ P=0 \text{~and~} J=0\\
		\{(i,X_i) \mid i\in \mathbb{T}\}, & \text{if} \  1 \le P \le \kappa(n) \text{~and~} J=0\\
		\mathfrak{C}, & \text{if} \ P> \kappa(n) \text{~or~} J=1,
	\end{cases}
\end{equation} 
where $\varnothing$ means that the channel is idle and $\mathfrak{C}$ means that the receiver observes a collision. There are two kinds of collision events. Collisions of the first type are called \textit{intrinsic} and are caused when the number of transmissions is above the network capacity, i.e., $P>\kappa(n)$. The second type of collisions are called \textit{extrinsic} and are caused when the jammer decides to block the channel, i.e., $J=1$.

Finally, for the measurement at the $i$-th sensor, the estimator uses a policy $\eta_i$ determined by
\begin{equation} \label{eq: estimatorn}
	\eta_i(y) =\begin{cases}
		x_i, & \text{if} \ \  P \le \kappa(n), \ i \in \mathbb{T}  \text{~and~} J=0 \\ 
		\hat{x}_{i0}, & \text{if}\ \ P \le \kappa(n), \ i \notin \mathbb{T}  \text{~and~} J=0  \\ 
		\hat{x}_{i1}, & \text{if}\ \ P> \kappa(n)
		\text{~or~} J=1,
	\end{cases}
\end{equation}
where $\hat{x}_{i0},\hat{x}_{i1} \in \mathbb{R}^m$ are the representation symbols used by the estimator when the $i$-th sensor's observation is not transmitted and when a collision occurs, respectively. Since the observations at all sensors are i.i.d., it is natural to assume that the sensors use the same transmission strategy $\gamma$. Similarly, the estimators use the same estimation strategy $\eta$, i.e., $ \hat{x}_{i0}=\hat{x}_0$ and $\hat{x}_{i1} = \hat{x}_1$, $i\in\{1,\cdots,n\}$. Let $\hat{x} \Equaldef  (\hat{x}_0,\hat{x}_1)$.  Using the law of total expectation and the mutual independence of $J$, $U_i$ and $\{U_\ell\}_{\ell \neq i}$, the objective function in \cref{eq:objectiven} can be expressed as
{\small
	\begin{multline} \label{eq3:objectiven}\mathcal{J}_n\big( (\gamma,\eta),\varphi \big) 
		= \frac{1}{n} \sum_{i=1}^n \bigg\{ \varphi \Big(\mathbf{E}\big[ (X_i-\hat{x}_{1})^2 \big] -d\Big) + c\Pr(U_i=1) \\
		+(1-\varphi)\bigg[\mathbf{E}\big[ (X_i-\hat{x}_{0})^2 \mid U_i=0\big] \mathbf{P}(U_i=0) \Pr\bigg(\sum_{\ell \neq i} U_\ell \le \kappa(n)\bigg)  \\
		+ \mathbf{E}\big[ (X_i-\hat{x}_{1})^2 \mid U_i=0\big] \Pr(U_i=0) \Pr\bigg(\sum_{\ell \neq i} U_\ell > \kappa(n)\bigg) \\
		+ \mathbf{E}\big[ (X_i-\hat{x}_{1})^2 \mid U_i=1\big] \Pr(U_i=1) \Pr\bigg(\sum_{\ell \neq i} U_\ell > \kappa(n)-1\bigg)\bigg]\bigg\}.
	\end{multline}}

Since $\{X_i\}_{i=1}^n$ is i.i.d., we may simplify the objective function in \cref{eq3:objectiven} as follows:
	\begin{multline} \label{eq4:objectiven}
		\mathcal{J}_n\big( (\gamma,\eta),\varphi \big) 
		= \varphi\Big(\mathbf{E}\big[ (X-\hat{x}_{1})^2 \big] -d\Big)  + c\Pr(U=1) 
		\\+(1-\varphi)\bigg[\mathbf{E}\Big[ (X-\hat{x}_{0})^2 \big(1-\gamma(X)\big)\Big] \mathcal{F}_{n-1,\kappa(n)}(\gamma)  \\
		 + \mathbf{E}\Big[ (X-\hat{x}_{1})^2 \big(1-\gamma(X)\big)\Big] \big(1-\mathcal{F}_{n-1,\kappa(n)}(\gamma)\big) \\ 
		+ \mathbf{E}\Big[ (X-\hat{x}_{1})^2 \gamma(X)\Big]  \big(1-\mathcal{F}_{n-1,\kappa(n)-1}(\gamma)\big)\bigg],				
	\end{multline}
where 
\begin{equation}
	\mathcal{F}_{n,\kappa}(\gamma) \Equaldef \sum_{m=0}^{\kappa}\binom{n}{m}\Pr(U=1)^m\big(1-\Pr(U=1)\big)^{n-m}.
\end{equation}

\subsection{Objective function in large-scale networks}

Taking the limit of $\mathcal{J}_n$ in \cref{eq4:objectiven} we can compute the objective function for large-scale networks. It is important to notice that we are constraining $\boldsymbol{\gamma}$ and $\boldsymbol{\eta}$ to be homogeneous strategy profiles. This can be justified by symmetry of the underlying probabilistic model. Additionally, the different sensing elements in large-scale systems are mass produced, and exhibit nearly identical characteristics. Therefore, it makes sense that the decision rules implemented by them are identical.

\vspace{5pt}


We will make use of the following variants of the standard Chernoff's inequality \cite[Theorems 4.4 and 4.5]{daglib0012859}.

\vspace{5pt}

\begin{lemma}[Chernoff's inequality] \label{lm: Chernoff_inequality} Consider a collection $\{U_{\ell}\}_{\ell=1}^n$ of independent Bernoulli variables with probability $p$. Let $S_n=\sum_{\ell=1}^{n} U_{\ell}$ and $\mu=\mathbf{E}[S_n]=np$, then
	\begin{itemize}
		\item[1)] $\mathbf{P}\big(S_n \ge (1+\delta) \mu \big) \le \exp(-\frac{\mu\delta^2}{2+\delta})$, for any $\delta>0$;
		\item[2)] $\mathbf{P}\big(S_n \le (1-\delta) \mu \big) \le \exp(-\frac{\mu\delta^2}{2})$, for any $0 < \delta <1$.
	\end{itemize}
\end{lemma}

\vspace{5pt}

\begin{lemma} \label{lm:limit} Suppose that $ \lim_{n \to \infty}  \kappa(n)/n=\bar{\kappa}$. Let $\{U_{\ell}\}_{\ell=1}^n$ be i.i.d.  Bernoulli variables with $\Pr(U_{\ell}=1)=p$ and $\Pr(U_{\ell}=0)=1-p$.  The following holds
\begin{equation}
\lim_{n \to \infty} \Pr \bigg( \sum_{\ell=1}^n U_\ell \leq \kappa(n)\bigg) = \begin{cases}1 & \text{if} \ \ p<\bar{\kappa} \\
0 & \text{if} \ \ p>\bar{\kappa}. \end{cases} 
\end{equation}
\end{lemma}

\vspace{5pt}

\begin{IEEEproof} 
Since $\lim_{n \to \infty} \kappa(n)/n =\bar{\kappa}$, then for any $\varepsilon>0$, there exists a natural number $N$ such that
	\begin{equation}
		\bigg|\frac{\kappa(n)}{n}-\bar{\kappa}\bigg|< \varepsilon, \  n \ge N.
	\end{equation}
\begin{itemize}
	\item[1)] For $p<\bar{\kappa}$, 
	fix $\varepsilon$ such that $\varepsilon\le (\bar{\kappa}-p)/2$. Then, there exists an $N$ such that
\begin{equation}
		\delta \Equaldef \frac{\kappa(n)}{np}-1 \in \Big(\frac{\bar{\kappa}-p}{2p}, \frac{3(\bar{\kappa}-p)}{2p}\Big).
	\end{equation}

	
	From Lemma \ref{lm: Chernoff_inequality}, we have
	\begin{multline}
		\mathbf{P}\bigg(\sum_{\ell=1}^nU_\ell > \kappa(n)\bigg)  =  \mathbf{P}\big(S_n > (1+\delta) \mu\big) \\
		 \le  \exp \Big(-\frac{\mu \delta^2}{2+\delta}\Big)  
		 \le  \exp \bigg(-\frac{np (\frac{\bar{\kappa}-p}{2p})^2}{2+\frac{3(\bar{\kappa}-p)}{2p}}\bigg) \\
		  =  \exp \bigg(-\frac{n (\bar{\kappa}-p)^2}{6\bar{\kappa}+2p}\bigg).
	\end{multline}
	
	Let $N' > \frac{6\bar{\kappa}+2p}{(\bar{\kappa}-p)^2} \ln (\frac{1}{\varepsilon})$, then for all
	$n \ge \max \{N,N'\}$, we have
	 	\begin{align}
	 	\mathbf{P}\bigg(\sum_{\ell=1}^nU_\ell > \kappa(n)\bigg) < \varepsilon.
	 \end{align}
 
 	Therefore, 
	
\begin{multline}
\lim_{n\rightarrow \infty}\mathbf{P}\bigg(\sum_{\ell=1}^nU_\ell > \kappa(n)\bigg) = 0 \\
\Rightarrow \lim_{n\rightarrow \infty}\mathbf{P}\bigg(\sum_{\ell=1}^nU_\ell \leq \kappa(n)\bigg) = 1.
\end{multline}


	\item[2)] For $p>\bar{\kappa}$, 
fix $\varepsilon$ such that $\varepsilon \le \min\{p-\bar{\kappa}, \bar{\kappa}\}/2$. Then, there exists an $N$ such that 
		\begin{equation}
		\delta \Equaldef 1-\frac{\kappa(n)}{np} \in \Big(\frac{p-\bar{\kappa}}{2p}, \min\Big\{\frac{3(p-\bar{\kappa})}{2p},\frac{2p-\bar{\kappa}}{2p}\Big\}\Big),
		\end{equation}	

		for all $n\geq N$. 
		From \cref{lm: Chernoff_inequality}, we have
	\begin{multline}
		\Pr \bigg( \sum_{\ell=1}^n U_\ell \leq \kappa(n)\bigg)  = \mathbf{P}\big(S_n \le (1-\delta) \mu\big) \\
		\le \exp \Big(-\frac{\mu \delta^2}{2}\Big)  
		\le \exp \bigg(-\frac{np (\frac{p-\bar{\kappa}}{2p})^2}{2}\bigg)\\
		= \exp \bigg(-\frac{n(p-\bar{\kappa})^2}{8p}\bigg).
	\end{multline}
	Let $N' > \frac{8p}{(p-\bar{\kappa})^2} \ln (\frac{1}{\varepsilon})$, then for all
	$n \ge \max \{N,N'\}$, we have
	\begin{align}
		\Pr \bigg( \sum_{\ell=1}^n U_\ell \leq \kappa(n)\bigg)< \varepsilon.
	\end{align}	
	 Therefore, 
	 \begin{equation}
	 \lim_{n \to \infty} \Pr \bigg( \sum_{\ell=1}^n U_\ell \leq \kappa(n)\bigg)=0.
\end{equation}
\end{itemize}
\end{IEEEproof}

	 

\begin{proposition} \label{prop:limit}
If $\lim_{n\rightarrow \infty}\kappa(n)/n = \bar{\kappa}$, then the following holds:
\begin{equation}
\lim_{n\rightarrow \infty}\mathcal{F}_{n-1,\kappa(n)}(\gamma) = \mathbf{1}\big(\Pr(U=1)\leq \bar{\kappa} \big) \ \ \mathrm{a.e.}
\end{equation}
and
\begin{equation}
\lim_{n\rightarrow \infty}\mathcal{F}_{n-1,\kappa(n)-1}(\gamma) = \mathbf{1}\big(\Pr(U=1)\leq \bar{\kappa} \big) \ \ \mathrm{a.e.}
\end{equation}
\end{proposition}

Therefore, in the asymptotic regime, we have 
\begin{multline} \label{eq1:objectiven_inf}
	\mathcal{J}_{\infty}  \big((\gamma,\eta),\varphi \big)
	= \varphi \big(\mathbf{E}\big[ (X-\hat{x}_{1})^2 \big] -d \big) + c\Pr(U=1)  \\
	+(1-\varphi)\Big[\mathbf{E}\big[ (X-\hat{x}_{0})^2 \mathbf{1}(U=0)\big] \mathbf{1}\big(\Pr(U=1)\leq \bar{\kappa} \big)   \\
	+ \mathbf{E}\big[ (X-\hat{x}_{1})^2\big] \mathbf{1}\big(\Pr(U=1)> \bar{\kappa} \big) \Big],
\end{multline}
which is equivalent to
\begin{multline}\label{eq:objectiven_inf3}
	\mathcal{J}_{\infty}  \big((\gamma,\eta),\varphi \big) = \varphi\big(\mathbf{E}\big[ (X-\hat{x}_{1})^2\big] - d \big) +c\Pr(U=1)  \\+
	 \begin{cases}		
		 (1-\varphi)\mathbf{E}\big[ (X-\hat{x}_{0})^2 \mathbf{1}(U=0)\big] , & \text{if~} \Pr(U=1) \leq \bar{\kappa} \\
		(1-\varphi)\mathbf{E}\big[ (X-\hat{x}_{1})^2 \big], &  \text{if~} \Pr(U=1) > \bar{\kappa}.
	\end{cases} 
\end{multline}


Since the coordinator can choose the value of $\Pr(U=1)$ by adjusting the transmission policy, the problem is equivalent to solving the following two problems and choosing the one with the smaller optimal value:
\begin{align} \label{opt-prb1}
	&\min_{\gamma, \eta} \max_{\varphi} \ \ \varphi\left(\mathbf{E}\big[ (X-\hat{x}_{1})^2 \big]- d \right)  + c\Pr(U=1) \nonumber\\
	& \qquad \qquad + (1-\varphi)\mathbf{E}\big[ (X-\hat{x}_{0})^2 \mathbf{1}(U=0)\big] \nonumber\\
	&\text{subject to~~}  \Pr(U=1) \le \bar{\kappa},
\end{align}
and
\begin{align} \label{opt-prb2}
	&\min_{\gamma,\eta} \max_{\varphi}  \ \ \varphi \left(\mathbf{E}\big[ (X-\hat{x}_{1})^2 \big]- d \right) + c\Pr(U=1) \nonumber\\
	& \qquad \qquad + (1-\varphi)\mathbf{E}\big[ (X-\hat{x}_{1})^2 \big] \nonumber\\
	&\text{subject to~~}  \Pr(U=1) > \bar{\kappa}.
\end{align}

\subsection{Characterization of saddle points solutions}

Let 
\begin{multline}
\mathcal{L}\big( (\gamma,\eta),\varphi \big) \Equaldef \varphi \left(\mathbf{E}[ (X-\hat{x}_{1})^2]- d \right) + c\Pr(U=1) \\ + (1-\varphi)\mathbf{E}\big[ (X-\hat{x}_{0})^2 \mathbf{1}(U=0)\big],
\end{multline}
with $\Pr(U=1) \le \bar{\kappa}$, and
\begin{multline}
\mathcal{U}\big( (\gamma,\eta),\varphi \big) = \varphi \left(\mathbf{E}\big[ (X-\hat{x}_{1})^2 \big]- d \right) + c\Pr(U=1) \\+ (1-\varphi)\mathbf{E}\big[ (X-\hat{x}_{1})^2 \big],
\end{multline}
with  $\Pr(U=1) > \bar{\kappa}$. The following result shows that the optimal objective function value is always obtained by solving \cref{opt-prb1}. 

\vspace{5pt}

\begin{proposition} \label{prop: fucntion_superiority_pro}
	Let $\gamma^\star,\eta^\star$ and $\varphi^\star$ be a saddle point of \cref{opt-prb1}. Then, we have 
\begin{align}\label{eq:ineq_saddle}
	\mathcal{L}\big( (\gamma^\star,\eta^\star), \varphi^\star \big) \le \mathcal{U}\big( (\gamma,\eta),\varphi \big), 
\end{align}
for all $\gamma$ such that $\Pr(U=1) > \bar{\kappa}$, and for all $\varphi\in [0,1]$.
\end{proposition}

\vspace{5pt}

\begin{IEEEproof}
For the two problems in \cref{opt-prb1,opt-prb2}, we have 
\begin{align}
\hat{x}_{1}^\star = \mu.
\end{align}

The solution of $\mathcal{U}( (\gamma,\eta),\varphi)$ is lower-bounded by setting $\Pr(U=1)=\bar{\kappa}$ and $\varphi=0$, i.e.,
\begin{equation} \label{opt-prb2-lower-bound}
\mathcal{U}( (\gamma,\eta),\varphi ) > \mathbf{E}\big[ (X-\mu)^2 \big]+ c \bar{\kappa}.
\end{equation}

If $\gamma^\star,\hat{x}_0^\star$ and $\varphi^\star$ is a saddle point of \cref{opt-prb1}, then the objective function in \cref{opt-prb1} satisfies 
\begin{align} \label{opt-prb1-upper-bound}
	\mathcal{L}\big( (\gamma^\star,\eta^\star), \varphi^\star \big)
	= &
 \varphi^\star \left(\mathbf{E}\big[ (X-\mu)^2 \big]- d \right) + c\Pr(U=1) \nonumber\\
	& \hspace{30pt} + (1-\varphi^\star)\mathbf{E}\big[ (X-\hat{x}_{0}^\star)^2 \mathbf{1}(U=0)\big] \nonumber \\
	\stackrel{(a)}{\le} & 
 \varphi^\star \left(\mathbf{E}\big[ (X-\mu)^2 \big]- d \right) + c\Pr(U=1) \nonumber\\
	&  \hspace{30pt}  + (1-\varphi^\star)\mathbf{E}\big[ (X-\mu)^2 \mathbf{1}(U=0)\big] \nonumber \\
	\stackrel{(b)}{\le} & 
 \varphi^\star \left(\mathbf{E}\big[ (X-\mu)^2 \big]- d \right) + c\Pr(U=1) \nonumber\\
	& \hspace{30pt} + (1-\varphi^\star)\mathbf{E}\big[ (X-\mu)^2\big] \nonumber \\
	{=} & \ \mathbf{E}\big[ (X-\mu)^2 \big]- d\varphi^\star + c\Pr(U=1) \nonumber\\
	\stackrel{(c)}{\le} & \ \mathbf{E}\big[ (X-\mu)^2 \big] + c\bar{\kappa}, 
\end{align}
where $(a)$ follows from the definition of saddle point equilibrium, $(b)$ follows from the inequality $\mathbf{E}\big[(X-\mu)^2 \mathbf{1}(U=0)\big] \le  \mathbf{E}\big[(X-\mu)^2\big]$ and $(c)$ holds due to $d\varphi^\star\ge 0$ and $\Pr(U=1) \le \bar{\kappa}$ for $\mathcal{L}\big( (\gamma,\eta), \varphi \big)$. Combining \cref{opt-prb2-lower-bound} and \cref{opt-prb1-upper-bound}, we obtain \cref{eq:ineq_saddle}.
\end{IEEEproof}

\vspace{5pt}

Therefore, it suffices to consider the constrained optimization problem in \cref{opt-prb1}. Define the Lagrangian function 
\begin{multline}
	L(\gamma, \eta, \varphi, \lambda) \Equaldef \varphi \big[\mathbf{E}\big[ (X-\hat{x}_1)^2 \big]-d \big]  + c\Pr(U=1) \\ + (1-\varphi)\mathbf{E}\big[ (X-\hat{x}_{0})^2 \mathbf{1}(U=0)\big] + \lambda \big(\Pr(U=1) - \bar{\kappa}\big),
\end{multline}
where $\lambda \ge 0$ is the dual variable associated with the inequality constraint $\Pr(U=1) - \bar{\kappa}\le 0$. 


\vspace{5pt}

\begin{proposition}[Optimality of threshold policies]\label{prop:threshold_n_inf} For a large-scale system under a proactive attack with a fixed jamming probability $\varphi \in[0,1]$, dual variable $\lambda$, and an arbitrary estimation policy $\eta$ indexed by representation symbols $\hat{x}=(\hat{x}_0,\hat{x}_1)\in \mathbb{R}^{2}$, the optimal transmission strategy is:
\begin{equation} \label{tran_policy_n_inf}
	\gamma_{\eta,\lambda,\varphi}^\star(x)= \mathbf{1}\big((1-\varphi)(x-\hat{x}_{0})^2 \ge c+\lambda \big).
\end{equation}
\end{proposition}

\vspace{5pt}

For fixed $\eta$, $\lambda$, and $\varphi$, \cref{prop:threshold_n_inf} implies that
\begin{multline}
	\tilde{L}\big(\eta, (\varphi, \lambda)\big) \Equaldef L(\gamma^\star_{\eta,\lambda,\varphi}, \eta, \varphi, \lambda) =  \varphi \big(\mathbf{E}\big[ (X-\hat{x}_1)^2 \big]-d \big) \\ + \mathbf{E}\Big[ \min \big\{(1-\varphi)(X-\hat{x}_{0})^2, c+\lambda \big\}\Big] - \lambda \bar{\kappa}.
\end{multline}


\begin{proposition}[Optimal estimator] \label{prop: optimal_estimator_n_inf} Let $X$ be a Gaussian random variable with mean $\mu$ and variance $\sigma^2$. The optimal estimator is
	\begin{equation} 
		\eta^\star (y) = \begin{cases}
			\mu, & \text{if} \ \ y \in \{\varnothing,\mathfrak{C}\} \\
			x, & \text{if} \ \ y = x.
		\end{cases}
	\end{equation}
\end{proposition}

Without loss of generality, set $\mu=0$. Then the Lagrangian function becomes
\begin{multline}
	\tilde{\tilde{L}}(\varphi, \lambda) \Equaldef \tilde{L}\big(\eta^\star, (\varphi, \lambda)\big) = \varphi \big(\mathbf{E}[ X^2 ]-d \big)  \\ + \mathbf{E}\Big[ \min \big\{(1-\varphi)X^2, c+\lambda \big\}\Big] - \lambda \bar{\kappa}.
\end{multline}

The optimal values $\varphi^\star$ and $\lambda^\star$ are coupled. Therefore, we must jointly maximize $\tilde{\tilde{L}}$ over $\varphi$ and $\lambda$. Let $l_{\lambda}(\bar{\kappa})$ denote the unique solution of 
\begin{align} \label{eq: l_lambda}
	2\int_{\sqrt{l_{\lambda}}}^{+\infty} f(x) \mathrm{d} x =\bar{\kappa}
\end{align}
and let $l_{\varphi}(d)$ denote the unique solution of
\begin{align} \label{eq: l_phi}
	2\int_{\sqrt{l_{\varphi}}}^{+\infty} x^2 f(x) \mathrm{d} x = d.
\end{align}

\vspace{5pt}

\begin{theorem}[Optimal jamming policy] \label{thm:jamming_Lagrange} For a given input pdf $f$, transmission cost $c$, jamming cost $d$, and asymptotic channel capacity $\bar{\kappa}$, the optimal jamming probability and its associated optimal Lagrange dual variable are:

\vspace{5pt}

\begin{enumerate}
\item If $\int_{\sqrt{c}}^{+\infty} f(x) \mathrm{d} x <\bar{\kappa}/2$ and $\int_{\sqrt{c}}^{+\infty} x^2 f(x) \mathrm{d} x <d/2$, then
\begin{equation}
	 (\varphi^\star,\lambda^\star)=(0,0);
\end{equation}
\item If $\int_{\sqrt{c}}^{+\infty} f(x) \mathrm{d} x \ge \bar{\kappa}/2$ and $\int_{\sqrt{c}}^{+\infty} x^2 f(x)\mathrm{d} x < d/2$, then
\begin{equation}
	(\varphi^\star,\lambda^\star)=(0,l_{\lambda}\big(\bar{\kappa})-c\big);
\end{equation}
\item If $\int_{\sqrt{c}}^{+\infty} f(x) \mathrm{d} x <\bar{\kappa}/2$ and $\int_{\sqrt{c}}^{+\infty} x^2 f(x) \mathrm{d} x \ge d/2$, then
\begin{equation}
(\varphi^\star,\lambda^\star)=\bigg(1-\frac{c}{l_{\varphi}(d)},0 \bigg);
\end{equation}
\item If $\int_{\sqrt{c}}^{+\infty} f(x) \mathrm{d} x \ge \bar{\kappa}/2$ and $\int_{\sqrt{c}}^{+\infty} x^2 f(x)\mathrm{d} x \ge d/2$, 
\begin{enumerate}[a.]
	\item[4.a)]  if $l_{\lambda}(\bar{\kappa})=l_{\varphi}(d)$, then 
	\begin{align}
		(\varphi^\star,\lambda^\star) \in \Bigg\{ (\varphi,\lambda) \in [0,1]\times\mathbb{R}_+ \ \bigg| \ \frac{c+\lambda}{1-\varphi}=l_{\lambda}(\kappa) \Bigg\};
	\end{align}
	\item[4.b)]  if $l_{\lambda}(\bar{\kappa})>l_{\varphi}(d)$, then
	\begin{equation}
		(\varphi^\star,\lambda^\star) = \big(0,l_{\lambda}(\bar{\kappa})-c\big);
	\end{equation}
	\item[4.c)]  if $l_{\varphi}(d)>l_{\lambda}(\bar{\kappa})$, then
	\begin{equation}
		(\varphi^\star,\lambda^\star) = \bigg(1-\frac{c}{l_{\varphi}(d)},0\bigg). 
	\end{equation}
\end{enumerate}
\end{enumerate}
\end{theorem}

\vspace{5pt}

\begin{IEEEproof}
The proof is in Appendix \ref{appendix:large_scale}.
\end{IEEEproof}

\vspace{5pt}

Next, we establish the existence of a saddle point equilibrium for \cref{opt-prb1}, i.e.,
\begin{equation}\label{eq:SPE_inf}
	\mathcal{L}\big( (\gamma^\star_{\eta^\star,\varphi^\star},\eta^\star),\varphi \big) \leq \mathcal{L}\big( (\gamma^\star_{\eta^\star,\varphi^\star},\eta^\star),\varphi^\star \big) \leq \mathcal{L}\big( (\gamma,\eta),\varphi^\star \big)
\end{equation}
for $\varphi\in [0,1]$ and $\gamma \in \{\gamma: \mathbb{R}\rightarrow [0,1] \mid \mathbf{P}\big(\gamma(X)=1\big)\le \bar{\kappa}\}$.
We will show that it suffices to show that the saddle point equilibrium of its Lagrangian function
\begin{multline}\label{eq:SPE_inf_L}
	L\big( (\gamma^\star,\eta^\star),(\varphi,\lambda^\star) \big) \leq L\big( (\gamma^\star,\eta^\star),(\varphi^\star,\lambda^\star) \big) \\ \leq L\big( (\gamma,\eta),(\varphi^\star,\lambda^\star)\big).
\end{multline}

\vspace{5pt}

\begin{proposition} \label{propo: saddle_same}
	Let $\big((\gamma^\star,\eta^\star),(\varphi^\star,\lambda^\star)\big)$ be a saddle point of $L\big((\gamma, \eta), (\varphi, \lambda)\big)$, then $\big((\gamma^\star,\eta^\star), \varphi^\star\big)$ is the saddle point of $\mathcal{L}\big((\gamma,\eta),\varphi\big)$.	
\end{proposition}

\vspace{5pt}

\begin{IEEEproof}
	Since $\big((\gamma^\star,\eta^\star),(\varphi^\star,\lambda^\star)\big)$ is a saddle point of $L((\gamma, \eta), (\varphi, \lambda))$ it must satisfy complementary slackness, i.e., 
	\begin{equation}
	\lambda^\star \Big(\mathbf{E}\big[\gamma^\star(X)\big] - \bar{\kappa}\Big)=0.
	\end{equation}
	Therefore, we always have 
	\begin{equation}
		L\big((\gamma^\star, \eta^\star), (\varphi^\star, \lambda^\star)\big)=\mathcal{L}\big( (\gamma^\star,\eta^\star),\varphi^\star \big).
	\end{equation}
	Since $\lambda \Big(\mathbf{E}\big[\gamma^\star(X)\big]-\bar{\kappa}\Big)\le 0$, we have 
	\begin{equation}
		L\big((\gamma, \eta), (\varphi^\star, \lambda^\star)\big) \le \mathcal{L}\big( (\gamma,\eta),\varphi^\star \big).
	\end{equation}
	When $\gamma=\gamma^\star$\footnote{Here, $\gamma^\star$ denotes the optimal transmission policy for given $\lambda^\star,\eta^\star,\varphi^\star$ as established in \cref{prop:threshold_n_inf}.}, the complementary slackness property is satisfied. 
	 Then \cref{thm:jamming_Lagrange} 
	 implies that
\begin{equation}
L\big( (\gamma^\star,\eta^\star),(\varphi,\lambda^\star)  \big)=\mathcal{L}\big( (\gamma^\star,\eta^\star),\varphi \big).
\end{equation} Therefore, 
	\begin{equation}
		\mathcal{L}\big( (\gamma^\star,\eta^\star),\varphi \big) \leq \mathcal{L}\big( (\gamma^\star,\eta^\star),\varphi^\star \big) \leq \mathcal{L}\big( (\gamma,\eta),\varphi^\star \big).
	\end{equation}
\end{IEEEproof}

Following the proof of \cref{thm:saddle point} and using \cref{propo: saddle_same}, we establish a saddle point equilibrium for large-scale networks.

\vspace{5pt}

\begin{theorem}[Saddle point equilibrium]\label{thm:saddle-inf} Given a  Gaussian source $X\sim \mathcal{N}(0,\sigma^2)$, communication and jamming costs $c,d\geq 0$, a saddle point strategy $(\gamma^\star,\eta^\star,\varphi^\star)$ for the remote estimation game with a proactive jammer over a large-scale network of capacity $\bar{\kappa}$ is given by:
	\begin{enumerate}
		\item If $\int_{\sqrt{c}}^{+\infty} f(x) \mathrm{d} x <\bar{\kappa}/2$ and $\int_{\sqrt{c}}^{+\infty} x^2 f(x) \mathrm{d} x <d/2$, then
	     \begin{align}\label{eq:saddle_inf_1}
			\gamma^\star(x) = \mathbf{1}(x^2 > c) \ \ \text{and} \ \ 
			\varphi^\star =0.
		\end{align}
		\item If $\int_{\sqrt{c}}^{+\infty} f(x) \mathrm{d} x \ge \bar{\kappa}/2$ and $\int_{\sqrt{c}}^{+\infty} x^2 f(x)\mathrm{d} x < d/2$, then
		\begin{align}\label{eq:saddle_inf_3}
			\gamma^\star(x) = \mathbf{1}\big(x^2 > l_{\lambda}(\bar{\kappa}) \big) \ \ \text{and} \ \
			\varphi^\star=0.
		\end{align}
			\item If $\int_{\sqrt{c}}^{+\infty} f(x) \mathrm{d} x <\bar{\kappa}/2$ and $\int_{\sqrt{c}}^{+\infty} x^2 f(x) \mathrm{d} x \ge d/2$, then
		\begin{align}\label{eq:saddle_inf_2}
			\gamma^\star(x) = \mathbf{1}\big(x^2 > l_{\varphi}(d) \big) \ \ \text{and} \ \
			\varphi^\star=1-\frac{c}{l_{\varphi}(d)}.
		\end{align}
		\item If $\int_{\sqrt{c}}^{+\infty} f(x) \mathrm{d} x \ge \bar{\kappa}/2$ and $\int_{\sqrt{c}}^{+\infty} x^2 f(x)\mathrm{d} x \ge d/2$, 
		\begin{enumerate}
			\item[4.a)]  if $l_{\lambda}(\bar{\kappa})=l_{\varphi}(d)$, then
					\begin{align}\label{eq:saddle_inf_4_1}
			\gamma^\star(x) = \mathbf{1}\big(x^2 > l_{\lambda}(\bar{\kappa}) \big) \ \ \text{and} \ \
			\varphi^\star \in \bigg[0,1-\frac{c}{l_{\varphi}(d)}\bigg];
			\end{align}
			\item[4.b)]  if $l_{\lambda}(\bar{\kappa})>l_{\varphi}(d)$, then
			\begin{align}\label{eq:saddle_inf_4_2}
				\gamma^\star(x) = \mathbf{1}\big(x^2 > l_{\lambda}(\bar{\kappa}) \big) \ \ \text{and} \ \
				\varphi^\star=0;
			\end{align}
			\item[4.c)]  if $l_{\varphi}(d)>l_{\lambda}(\bar{\kappa})$, then
				\begin{align}\label{eq:saddle_inf_4_3}
					\gamma^\star(x) = \mathbf{1}\big(x^2 > l_{\varphi}(d) \big) \ \ \text{and} \ \
					\varphi^\star=1-\frac{c}{l_{\varphi}(d)}.
				\end{align}
		\end{enumerate}
	\end{enumerate}
In all cases, the estimation policy is:
\begin{equation}
\eta^\star (y) = \begin{cases}
				0, & \text{if} \ \ y \in \{\varnothing,\mathfrak{C}\} \\
				x, & \text{if} \ \ y = x.
			\end{cases}
\end{equation}
\end{theorem}

\begin{table} [!t]
	\caption{Saddle point equilibrium for different problem parameters}
	\begin{center}
		\begin{tabular}{c c c c c c}  
			\hline
			$d$ &$\bar{\kappa}$ & $\gamma^\star$& $\mathbf{P}(\gamma^\star(X)=1)$  & $\varphi^\star$ & $\lambda^\star$ \\
			\hline\hline
			$0.25$ &   $0.25$  &  $\mathbf{1}(x^2 > 4.11)$   &  $0.04$ & $0.76$ &$0$ \\ 
			$0.25$ &   $0.50$  &  $\mathbf{1}(x^2 > 4.11)$   &  $0.04$ & $0.76$ &$0$ \\ 
			$0.25$ &   $0.75$  &  $\mathbf{1}(x^2 > 4.11)$   &  $0.04$ & $0.76$ &$0$\\
			$0.50$ &   $0.25$  &  $\mathbf{1}(x^2 > 2.37)$   &  $0.12$ & $0.58$ &$0$\\
			$0.50$ &   $0.50$  &  $\mathbf{1}(x^2 > 2.37)$   &  $0.12$ & $0.58$ &$0$\\
			$0.50$ &   $0.75$  &  $\mathbf{1}(x^2 > 2.37)$   &  $0.12$ & $0.58$ &$0$\\
			$0.75$ &   $0.25$  &  $\mathbf{1}(x^2 > 1.32)$   &  $0.25$ & $0$    &$0.32$\\
			$0.75$ &   $0.50$  &  $\mathbf{1}(x^2 > 1.21)$   &  $0.27$ & $0.18$ &$0$  \\
			$0.75$ &   $0.75$  & $\mathbf{1}(x^2 > 1.21)$    &  $0.27$ & $0.18$ &$0$  \\
			$1.00$ &   $0.25$  &  $\mathbf{1}(x^2 > 1.32)$   &  $0.25$ & $0$    &$0.32$\\
			$1.00$ &   $0.50$  &  $\mathbf{1}(x^2 > 1.00)$   &  $0.32$ & $0$    &$0$ \\
			$1.00$ &   $0.75$  &  $\mathbf{1}(x^2 > 1.00)$   &  $0.32$ & $0$    &$0$ \\
			\hline
		\end{tabular} \label{table: FNE pairs}
	\end{center}
\end{table}

\subsection{Numerical results}

Based on \cref{thm:saddle-inf}, the following numerical results provide some insights on the optimal transmission strategy and optimal jamming strategy.  Table \ref{table: FNE pairs} shows the saddle point equilibrium under different parameters, where  $X \sim \mathcal{N}(0,1)$ and $c=1$.  For example, let $d=1$ and $\bar{\kappa}=0.25$. Since 
\begin{multline}
	\int_{\sqrt{c}}^{+\infty} f(x) \mathrm{d} x =0.16 > \bar{\kappa}/2, \ \ \text{and} \\
	\int_{\sqrt{c}}^{+\infty} x^2 f(x) \mathrm{d} x  = 0.40 < d/2,
\end{multline}
we have $\gamma^\star=\mathbf{1}(x^2 > 1.32)$ and $\varphi^\star=0$. Considering $d=0.25$ and $\bar{\kappa}=0.25$, we have
\begin{multline}
	\int_{\sqrt{c}}^{+\infty} f(x) \mathrm{d} x =0.16 > \bar{\kappa}/2, \ \ \text{and} \\
	\int_{\sqrt{c}}^{+\infty} x^2 f(x) \mathrm{d} x  = 0.40 > d/2,
\end{multline}
and $l_{\lambda}(\bar{\kappa})=1.32 < l_{\varphi}(d)=4.11$. So the optimal strategies are $\gamma^\star=\mathbf{1}(x^2 > 4.11)$ and $\varphi^\star=0.76$. Notice that the complementary slackness property is always satisfied, i.e., $\lambda^\star (\mathbf{P}(\gamma^\star(X)=1)-\bar{\kappa})=0$. Moreover, in the saddle point equilibrium of \cref{thm:saddle-inf}, there is a sharp transition in the optimal jamming probability from zero to nonzero, which directly depends on the jamming cost. However, the structure of the transmission and estimation policies remain unchanged. In particular, the optimal transmission threshold policy is always symmetric.

\section{Concluding remarks and future work}


Building upon the pioneering model introduced by Gupta et al. in \cite{gupta2012dynamic,gupta2016}, we have considered a remote estimation game with asymmetric information involving transmitters, receivers and a jammer. While most of the literature focuses on jamming in the network and the physical layer of the communication protocol stack, our work focuses on the medium access control layer. To address the complicated problem originated by the fact that the problem has a non-classical information structure, we adopt a coordinator approach, which leads to a tractable framework based on a zero-sum game between coordinator and the jammer. We have obtained several results on the saddle point equilibria for many cases of interest, and extended the result for large scale networks, which provide insights in the design of massive IoT deployments for many modern applications such as smart farming, Industry 4.0, and robotic swarms.

There are many interesting directions for future work. The most prominent ones are related to learning. In this work, we have assumed that the probability density function of the observations are common knowledge. However, this assumption is never realistic in practice. The design of real systems is data-driven which leads to issues related to the stability, robustness and performance bounds when the probabilistic model is not known a priori and is learned from data samples. For example, the sample complexity of our system is a largely unexplored issue with only a few related results reported in \cite{vasconcelos2022learning}. Additionally, all of our results assume that the jamming and communication costs are available to the coordinator and the jammer, which is also a contrived assumption. If the costs are private information, the game may no longer be zero-sum. Moreover, these parameters may need to be learned from repeated play. In such case, it would be interesting to developed a theory that characterize the rate of regret in online learning in this more realistic scenario.

\bibliographystyle{IEEEtran}

\bibliography{IEEEabrv,refs}

\begin{thebibliography}{10}
\providecommand{\url}[1]{#1}
\csname url@samestyle\endcsname
\providecommand{\newblock}{\relax}
\providecommand{\bibinfo}[2]{#2}
\providecommand{\BIBentrySTDinterwordspacing}{\spaceskip=0pt\relax}
\providecommand{\BIBentryALTinterwordstretchfactor}{4}
\providecommand{\BIBentryALTinterwordspacing}{\spaceskip=\fontdimen2\font plus
\BIBentryALTinterwordstretchfactor\fontdimen3\font minus
  \fontdimen4\font\relax}
\providecommand{\BIBforeignlanguage}[2]{{%
\expandafter\ifx\csname l@#1\endcsname\relax
\typeout{** WARNING: IEEEtran.bst: No hyphenation pattern has been}%
\typeout{** loaded for the language `#1'. Using the pattern for}%
\typeout{** the default language instead.}%
\else
\language=\csname l@#1\endcsname
\fi
#2}}
\providecommand{\BIBdecl}{\relax}
\BIBdecl

\bibitem{Vasconcelos:2020}
M.~M. Vasconcelos and N.~C. Martins, ``A survey on remote estimation
  problems,'' \emph{Principles of Cyber-Physical Systems: An Interdisciplinary
  Approach}, pp. 81--103, 2020.

\bibitem{xia2016networked}
M.~Xia, V.~Gupta, and P.~J. Antsaklis, ``Networked state estimation over a
  shared communication medium,'' \emph{IEEE Transactions on Automatic Control},
  vol.~62, no.~4, pp. 1729--1741, 2016.

\bibitem{Kang:2023}
S.~Kang, A.~Eryilmaz, and N.~B. Shroff, ``Remote tracking of distributed
  dynamic sources over a random access channel with one-bit updates,''
  \emph{IEEE Transactions on Network Science and Engineering}, pp. 1--11, 2023.

\bibitem{Raza2017low}
U.~Raza, P.~Kulkarni, and M.~Sooriyabandara, ``Low power wide area networks: An
  overview,'' \emph{IEEE Communications Surveys \& Tutorials}, vol.~19, no.~2,
  pp. 855--873, 2017.

\bibitem{basar1983gaussian}
T.~Basar, ``The {Gaussian} test channel with an intelligent jammer,''
  \emph{IEEE Transactions on Information Theory}, vol.~29, no.~1, pp. 152--157,
  1983.

\bibitem{mcdonald2019two}
C.~McDonald, F.~Alajaji, and S.~Y{\"u}ksel, ``Two-way {G}aussian networks with
  a jammer and decentralized control,'' \emph{IEEE Transactions on Control of
  Network Systems}, vol.~7, no.~1, pp. 446--457, 2019.

\bibitem{Akyol2017Info}
E.~Akyol, C.~Langbort, and T.~Ba{\c s}ar, ``Information-theoretic approach to
  strategic communication as a hierarchical game,'' \emph{Proceedings of the
  IEEE}, vol. 105, no.~2, pp. 205--218, 2017.

\bibitem{Akyol2019Optimal}
E.~Akyol, ``On optimal jamming in strategic communication,'' in \emph{2019 IEEE
  Information Theory Workshop (ITW)}, 2019, pp. 1--5.

\bibitem{Gao2019Communication}
X.~{Gao}, E.~{Akyol}, and T.~{Basar}, ``Communication scheduling and remote
  estimation with adversarial intervention,'' \emph{IEEE/CAA Journal of
  Automatica Sinica}, vol.~6, no.~1, pp. 32--44, 2019.

\bibitem{Shafiee2009Mutual}
S.~Shafiee and S.~Ulukus, ``Mutual information games in multiuser channels with
  correlated jamming,'' \emph{IEEE Transactions on Information Theory},
  vol.~55, no.~10, pp. 4598--4607, 2009.

\bibitem{ray2006optimal}
S.~Ray, P.~Moulin, and M.~Medard, ``On optimal signaling and jamming strategies
  in wideband fading channels,'' in \emph{2006 IEEE 7th Workshop on Signal
  Processing Advances in Wireless Communications}.\hskip 1em plus 0.5em minus
  0.4em\relax IEEE, 2006, pp. 1--5.

\bibitem{Altman:2011}
E.~Altman, K.~Avrachenkov, and A.~Garnaev, ``Jamming in wireless networks under
  uncertainty,'' \emph{Mobile Networks and Applications}, vol.~16, no.~2, pp.
  246--254, 2011.

\bibitem{aziz2020resilience}
F.~M. Aziz, L.~Li, J.~S. Shamma, and G.~L. St{\"u}ber, ``Resilience of {LTE}
  {eNode B} against smart jammer in infinite-horizon asymmetric repeated
  zero-sum game,'' \emph{Physical Communication}, vol.~39, 2020.

\bibitem{li2015jamming}
Y.~Li, L.~Shi, P.~Cheng, J.~Chen, and D.~E. Quevedo, ``Jamming attacks on
  remote state estimation in cyber-physical systems: A game-theoretic
  approach,'' \emph{IEEE Transactions on Automatic Control}, vol.~60, no.~10,
  pp. 2831--2836, 2015.

\bibitem{li2016sinr}
Y.~Li, D.~E. Quevedo, S.~Dey, and L.~Shi, ``{SINR}-based {DoS} attack on remote
  state estimation: A game-theoretic approach,'' \emph{IEEE Transactions on
  Control of Network Systems}, vol.~4, no.~3, pp. 632--642, 2016.

\bibitem{ding2017stochastic}
K.~Ding, S.~Dey, D.~E. Quevedo, and L.~Shi, ``Stochastic game in remote
  estimation under dos attacks,'' \emph{IEEE control systems letters}, vol.~1,
  no.~1, pp. 146--151, 2017.

\bibitem{Wu:2017}
Y.~Wu, Y.~Li, and L.~Shi, ``A game-theoretic approach to remote state
  estimation in presence of a {DoS} attacker,'' \emph{IFAC-PapersOnLine},
  vol.~50, no.~1, pp. 2595--2600, 2017.

\bibitem{ding2018attacks}
K.~Ding, X.~Ren, D.~E. Quevedo, S.~Dey, and L.~Shi, ``{DoS} attacks on remote
  state estimation with asymmetric information,'' \emph{IEEE Transactions on
  Control of Network Systems}, vol.~6, no.~2, pp. 653--666, 2018.

\bibitem{ding2017multi}
K.~Ding, Y.~Li, D.~E. Quevedo, S.~Dey, and L.~Shi, ``A multi-channel
  transmission schedule for remote state estimation under {DoS} attacks,''
  \emph{Automatica}, vol.~78, pp. 194--201, 2017.

\bibitem{Feng:2021}
Y.~Feng, Y.~Shou, and X.~Yu, ``Jamming on remote estimation over wireless links
  under faded uncertainty: A {Stackelberg} game approach,'' \emph{IEEE
  Transactions on Circuits and Systems II: Express Briefs}, vol.~68, no.~7, pp.
  2593--2597, 2021.

\bibitem{peng2017optimal}
L.~Peng, L.~Shi, X.~Cao, and C.~Sun, ``Optimal attack energy allocation against
  remote state estimation,'' \emph{IEEE Transactions on Automatic Control},
  vol.~63, no.~7, pp. 2199--2205, 2017.

\bibitem{gupta2012dynamic}
A.~Gupta, A.~Nayyar, C.~Langbort, and T.~Ba{\c{s}}ar, ``A dynamic
  transmitter-jammer game with asymmetric information,'' in \emph{51st
  Conference on Decision and Control (CDC)}.\hskip 1em plus 0.5em minus
  0.4em\relax IEEE, 2012, pp. 6477--6482.

\bibitem{gupta2016}
A.~Gupta, C.~Langbort, and T.~Ba{\c{s}}ar, ``Dynamic games with asymmetric
  information and resource constrained players with applications to security of
  cyberphysical systems,'' \emph{IEEE Transactions on Control of Network
  Systems}, vol.~4, no.~1, pp. 71--81, 2016.

\bibitem{vasconcelos:2017a}
M.~M. Vasconcelos and N.~C. Martins, ``Optimal estimation over the collision
  channel,'' \emph{IEEE Transactions on Automatic Control}, vol.~62, no.~1, pp.
  321--336, January 2017.

\bibitem{vasconcelos2018optimal}
------, ``Optimal remote estimation of discrete random variables over the
  collision channel,'' \emph{IEEE Transactions on Automatic Control}, vol.~64,
  no.~4, pp. 1519--1534, 2019.

\bibitem{Nayyar2014}
A.~Nayyar, A.~Mahajan, and D.~Teneketzis, \emph{The Common-Information Approach
  to Decentralized Stochastic Control}.\hskip 1em plus 0.5em minus 0.4em\relax
  Springer International Publishing, 2014, pp. 123--156.

\bibitem{zhang2022robust}
X.~Zhang and M.~M. Vasconcelos, ``Robust remote estimation over the collision
  channel in the presence of an intelligent jammer,'' in \emph{2022 IEEE 61st
  Conference on Decision and Control (CDC)}.\hskip 1em plus 0.5em minus
  0.4em\relax IEEE, 2022, pp. 5472--5479.

\bibitem{Sanjari:2021}
S.~Sanjari and S.~Y{\"u}ksel, ``Optimal solutions to infinite-player stochastic
  teams and mean-field teams,'' \emph{IEEE Transactions on Automatic Control},
  vol.~66, no.~3, pp. 1071--1086, 2021.

\bibitem{Hespanha:2017}
J.~P. Hespanha, \emph{Noncooperative game theory: An introduction for engineers
  and computer scientists}.\hskip 1em plus 0.5em minus 0.4em\relax Princeton
  University Press, 2017.

\bibitem{Ostrovskii:2021}
D.~M. Ostrovskii, A.~Lowy, and M.~Razaviyayn, ``Efficient search of first-order
  {Nash}-equilibria in nonconvex-concave smooth min-max problems,''
  \emph{{SIAM} Journal on Optimization}, vol.~31, no.~4, pp. 2508--2538, 2021.

\bibitem{Nouiehed:2019}
M.~Nouiehed, M.~Sanjabi, T.~Huang, J.~D. Lee, and M.~Razaviyayn, ``Solving a
  class of non-convex min-max games using iterative first order methods,''
  \emph{Advances in Neural Information Processing Systems}, vol.~32, 2019.

\bibitem{Facchinei:2003}
F.~Facchinei and J.-S. Pang, \emph{Finite-dimensional variational inequalities
  and complementarity problems}.\hskip 1em plus 0.5em minus 0.4em\relax
  Springer, 2003.

\bibitem{Lin:2020}
T.~Lin, C.~Jin, and M.~Jordan, ``On gradient descent ascent for
  nonconvex-concave minimax problems,'' in \emph{International Conference on
  Machine Learning}.\hskip 1em plus 0.5em minus 0.4em\relax PMLR, 2020, pp.
  6083--6093.

\bibitem{Yuille:2003}
A.~L. Yuille and A.~Rangarajan, ``The concave-convex procedure,'' \emph{Neural
  computation}, vol.~15, no.~4, pp. 915--936, 2003.

\bibitem{Lipp:2016}
T.~Lipp and S.~Boyd, ``Variations and extension of the convex-concave
  procedure,'' \emph{Optimization and Engineering}, vol.~17, no.~2, pp.
  263--287, 2016.

\bibitem{daglib0012859}
M.~Mitzenmacher and E.~Upfal, \emph{Probability and Computing: Randomized
  Algorithms and Probabilistic Analysis.}\hskip 1em plus 0.5em minus
  0.4em\relax Cambridge University Press, 2005.

\bibitem{vasconcelos2022learning}
M.~M. Vasconcelos, ``Learning distributed channel access policies for networked
  estimation: data-driven optimization in the mean-field regime,'' in
  \emph{Learning for Dynamics and Control Conference}.\hskip 1em plus 0.5em
  minus 0.4em\relax PMLR, 2022, pp. 702--712.

\end{thebibliography}

\appendices

\section{Extension to the vector case} \label{appendix: p2p_vector}
In this section, we extend the sensor's measurement $X$ from scalars to vectors. The first step is to consider diagonal covariance matrices. Let ${X} \in \mathbb{R}^m$ with $m\ge 1$. The objective function is 
\begin{equation}\label{eq:objective_n_m}
	\mathcal{J}\big( (\gamma,\eta),\varphi \big) = \mathbf{E}\big[ \|{X}-\hat{{X}}\|^2 \big] + c\Pr(U=1) - d\Pr(J=1),
\end{equation}
and the estimator policy is given by \cref{eq:estimator1}.
Denote $\hat{{x}}=(\hat{{x}}_0,\hat{{x}}_1)$. The observation ${X}$ is a multivariate Gaussian distribution 
with mean ${\mu}$ and diagonal covariance matrix with ${\Sigma}=\rm{diag}(\sigma_1^2,\cdots,\sigma_m^2)$.

%
%
%

\vspace{5pt}
\begin{theorem}\label{thm: optimal_estimator_n_m}
	Let ${X}$ follow multivariate Gaussian distribution with expectation $ \mu$ and diagonal covariance matrix  $ \Sigma$,  then 
	\begin{equation}
		\eta^\star ({y}) = \begin{cases}
			{\mu}, & \text{if} \ \ {y} \in \{\varnothing,\mathfrak{C}\} \\
			{x}, & \text{if} \ \ {y} = {x}.
		\end{cases}
	\end{equation}
\end{theorem}
\begin{IEEEproof} It is immediate that ${x}^\star_1={\mu}$. Next, we will show ${x}^\star_0={\mu}$. Define 
	\begin{equation}
		\mathcal{H}(\hat{{x}}_0) = \mathbf{E}\Big[\min\big\{(1-\varphi)\|{X}-\hat{{x}}_0\|^2, c \big\} \Big].
	\end{equation}
Note that $\nabla_{\hat{ x}_0} \mathcal{H}(\hat{{x}}_0)= \nabla_{\hat{ x}_0} \mathcal{J}(\hat{{x}}_0,\hat{{x}}_1)$. By rewriting $\mathcal{H}(\hat{{x}}_0)$ as the integral form, we obtain
\begin{align}
	\mathcal{H}(\hat{{x}}_0)= &\int_{-\infty}^{+\infty} \ldots  \int_{-\infty}^{+\infty} \min\big\{(1-\varphi)\|{x}-\hat{{x}}_0\|^2, c \big\}\nonumber\\
	&\cdot \frac{1}{\sqrt{(2\pi)^m|{\Sigma}|}} \exp\left( -\frac{1}{2}({x}-{\mu})^\mathsf{T} {\Sigma}^{-1} ({x}-{\mu}) \right) \mathrm{d} {x}\\
	=&\int_{-\infty}^{+\infty} \ldots  \int_{-\infty}^{+\infty} \min\big\{(1-\varphi) \|{z}\|^2, c \big\} \frac{1}{\sqrt{(2\pi)^m|{\Sigma}|}} \nonumber\\
	&\cdot  \exp\left( -\frac{1}{2}({z} + \hat{{x}}_0-{\mu})^\mathsf{T} {\Sigma}^{-1} ({z} + \hat{{x}}_0 -{\mu}) \right) \mathrm{d} {z}.
\end{align}

Taking derivative with respect to $\hat{{x}}_0$, we get
\begin{align}
	&\nabla_{\hat{ x}_0} \mathcal{J}(\hat{{x}}_0,\hat{{x}}_1)=\nabla_{\hat{ x}_0} \mathcal{H}(\hat{{x}}_0) \nonumber\\
	=&\int_{-\infty}^{+\infty} \ldots  \int_{-\infty}^{+\infty} \min\big\{(1-\varphi) \|{z}\|^2, c \big\} \frac{1}{\sqrt{(2\pi)^m|{\Sigma}|}} \nonumber\\
	&\cdot  \exp\left( -\frac{1}{2} ({z} + \hat{{x}}_0-{\mu})^\mathsf{T} {\Sigma}^{-1} ({z} + \hat{{x}}_0 -{\mu}) \right)  \nonumber\\
	&\cdot \left[- {\Sigma}^{-1} ({z} + \hat{{x}}_0 -{\mu}) \right]  \mathrm{d} {z}.
\end{align}

Substituting ${z} + \hat{{x}}_0 -{\mu}$ with $ v$ obtains
\begin{align}
	&\nabla_{\hat{ x}_0} \mathcal{J}(\hat{{x}}_0,\hat{{x}}_1)\nonumber\\
	=&\int_{-\infty}^{+\infty} \ldots  \int_{-\infty}^{+\infty} \min\big\{(1-\varphi) \|{v}-(\hat{{x}}_0 -{\mu})\|^2, c \big\}  \nonumber\\
	&\cdot \frac{1}{\sqrt{(2\pi)^m|{\Sigma}|}} \exp\left( -\frac{1}{2} {v}^\mathsf{T}  {\Sigma}^{-1} {v} \right) \left[- {\Sigma}^{-1} {v} \right]  \mathrm{d} {v}.
\end{align}

Since $ \Sigma$ is a diagonal covariance matrix, the $i$-th entry of $\nabla_{\hat{ x}_0} \mathcal{J}(\hat{{x}}_0,\hat{{x}}_1)$ can be represented as
{\small
\begin{multline} \label{entry-wise-derivative}
	[\nabla_{\hat{ x}_0}  \mathcal{J}(\hat{{x}}_0,\hat{{x}}_1)]_i = 
	\int_{-\infty}^{+\infty} \ldots  \int_{-\infty}^{+\infty} \min\big\{(1-\varphi) \|{v}-(\hat{{x}}_0 -{\mu})\|^2, c \big\}  \\
	\cdot \frac{1}{\sqrt{(2\pi)^m|{\Sigma}|}} \exp\bigg(-\sum_{j=1}^m \frac{v_j^2}{2\sigma_j^2}\bigg) \left[-\frac{v_i}{\sigma_i^2} \right]  \mathrm{d} {v}.
\end{multline}}

We rewrite  $\min\big\{(1-\varphi) \|{v}-(\hat{{x}}_0 -{\mu})\|^2, c \big\}$ as
{\small
\begin{multline}
	\min\big\{(1-\varphi) \|{v}-(\hat{{x}}_0 -{\mu})\|^2, c \big\} =
	 	\min\big\{(1-\varphi) [v_i-(\hat{x}_{0i} -\mu_i)]^2, \\
	  c -(1-\varphi) \sum_{j\neq i}^{m} [v_j-(\hat{x}_{0j} -\mu_j)]^2\big\} 
	+ (1-\varphi) \sum_{j\neq i}^{m} [v_j-(\hat{x}_{0j} -\mu_j)]^2.
\end{multline}}
Using the fact that the inner function is odd with respect to $v_i$, we have
\begin{align} \label{eq: v_-i=>0}
&\int_{-\infty}^{+\infty} \ldots  \int_{-\infty}^{+\infty} (1-\varphi) \sum_{-i} [v_j-(\hat{x}_{0j} -\mu_j)]^2  \nonumber\\
	&\cdot \frac{1}{\sqrt{(2\pi)^m|{\Sigma}|}} \exp\bigg(-\sum_{j=1}^m \frac{v_j^2}{2\sigma_j^2}\bigg) \left[-\frac{v_i}{\sigma_i^2} \right]  \mathrm{d} {v}=0,
\end{align}

Define ${v}_{-i} \in \mathbb{R}^{m-1}$ as the vector that removes $v_i$ from ${v}$. By using \cref{eq: v_-i=>0}, \cref{entry-wise-derivative} can be reformulated as 
\begin{align}
[\nabla_{\hat{x}_0}  \mathcal{J}(\hat{{x}}_0,\hat{{x}}_1)]_i 
=\int_{-\infty}^{+\infty} \ldots  \int_{-\infty}^{+\infty} h({v}_{-i}) g({v}_{-i}) \mathrm{d} {v}_{-i},
\end{align}
where 
\begin{align}
 h({v}_{-i}) &= \frac{1}{\sqrt{(2\pi)^m|{\Sigma}|}} \exp\bigg( -\sum_{j \neq i}^m \frac{v_j^2}{2\sigma_j^2}\bigg), \\
 g({v}_{-i}) &= \int_{-\infty}^{+\infty} \min\big\{(1-\varphi) [v_i-(\hat{x}_{0i} -\mu_i)]^2, \nonumber\\
 & \qquad \qquad  c -(1-\varphi) \sum_{j\neq i}^{m} [v_j-(\hat{x}_{0j}-\mu_j)]^2 \big\} \nonumber\\
 & \qquad \qquad \qquad \quad \cdot \exp\bigg(- \frac{v_i^2}{2\sigma_i^2}\bigg) \left[-\frac{v_i}{\sigma_i^2} \right] \mathrm{d} v_i.
\end{align}

Note that $h({v}_{-i})$ is positive for all ${v}_{-i}$. Define $\mathcal{E}= \{{v}_{-i} |(1-\varphi) \sum_{j\neq i}^m [v_j-(\hat{x}_{0j}-\mu_j)]^2<c\}$. For ${v}_{-i} \notin \mathcal{E}$, we have
\begin{align}  \label{entry-wise-derivative_1}
	g({v}_{-i}) &= \int_{-\infty}^{+\infty} \big[ c -(1-\varphi)  \sum_{j\neq i}^{m} [v_j-(\hat{x}_{0j}-\mu_j)]^2 \big] \nonumber\\
	& \qquad \qquad \quad \cdot \exp\bigg(- \frac{v_i^2}{2\sigma_i^2}\bigg) \left[-\frac{v_i}{\sigma_i^2} \right] \mathrm{d} v_i=0,
\end{align}
where the first equality uses $(1-\varphi) [v_i-(\hat{x}_{0i} -\mu_i)]^2 \ge 0$ and the second equality uses the oddness of inner function. For ${v}_{-i} \in \mathcal{E}$, we obtain the sign of  $g({v}_{-i})$ in three cases:
\begin{enumerate}
\item $\hat{x}_{0i}=\mu_i$. The oddness of the inner function implies that $g({v}_{-i})=0$.
\item $\hat{x}_{0i}>\mu_i$. A similar technique with the proof of \cref{thm: optimal_estimator_n_1} implies that $g({v}_{-i})>0$.
\item $\hat{x}_{0i}<\mu_i$. A similar technique with the proof of \cref{thm: optimal_estimator_n_1} implies that $g({v}_{-i})<0$.
\end{enumerate}

Finally, we analyze the sign of $[\nabla_{\hat{ x}_0}  \mathcal{J}(\hat{{x}}_0,\hat{{x}}_1)]_i$, which is 
\begin{multline}
	[\nabla_{\hat{ x}_0}  \mathcal{J}(\hat{{x}}_0,\hat{{x}}_1)]_i 
	=\int_{{v}_{-i} \in \mathcal{E}}  h({v}_{-i}) g({v}_{-i}) \mathrm{d} {v}_{-i} \\
	+ \int_{{v}_{-i} \notin \mathcal{E}}  h({v}_{-i}) g({v}_{-i}) \mathrm{d} {v}_{-i} 
	=\int_{{v}_{-i} \in \mathcal{E}}  h({v}_{-i}) g({v}_{-i}) \mathrm{d} {v}_{-i},
\end{multline}
where the last equality uses \cref{entry-wise-derivative_1}.  Together with the sign of $g({v}_{-i})$ for ${v}_{-i} \in \mathcal{E}$, we obtain 
\begin{equation}
	\begin{cases}
		[\nabla_{\hat{ x}_0}  \mathcal{J}(\hat{{x}}_0,\hat{{x}}_1)]_i < 0 & \text{if} \ \ \hat{x}_{0i}<\mu_i \\ 
		[\nabla_{\hat{ x}_0}  \mathcal{J}(\hat{{x}}_0,\hat{{x}}_1)]_i = 0 & \text{if}\ \ \hat{x}_{0i}=\mu_i\\ 
		[\nabla_{\hat{ x}_0}  \mathcal{J}(\hat{{x}}_0,\hat{{x}}_1)]_i > 0 & \text{if}\ \ \hat{x}_{0i}>\mu_i.
	\end{cases}
\end{equation}
Therefore, $\hat{{x}}_{0}={\mu}$ minimizes $\mathcal{J}(\hat{{x}}_0,\hat{{x}}_1^\star)$.
\end{IEEEproof}

{\color{black}
Therefore, the  general objective function  of \cref{eq:jammer_obj} for the jammer is formulated as
\begin{multline}\label{eq:jammer_obj_m}
	\mathcal{J}\big((\gamma_{\eta^\star,\varphi}^\star,\eta^\star),\varphi\big) = \mathbf{E}\bigg[\min\Big\{(1-\varphi) \|X\|^2, c \Big\} \bigg] \\ + \varphi\Big(\mathbf{E}\big[\|X\|^2\big] -d \Big) .
\end{multline}	
	
Define $\tilde{\varphi}$ as
\begin{equation} \label{eq: J_d_beta_tilde_vector}
	\tilde{\varphi} \Equaldef \bigg\{ \varphi \in [0,1) \ \Big| \  \mathbf{E}\Big[ \|X\|^2 \textbf{1}\big((1-\varphi)\|X\|^2 > c\big) \Big] =d\bigg\}.
\end{equation}

\begin{theorem} \label{thm: optimal_jam_prob_m} 
	Let ${X}$ be a multivariate Gaussian random vector with mean $\mu$ and diagonal covariance matrix  $ \Sigma$. The optimal jamming probability when the  transmission policy has the structure as in \cref{prop:threshold_n_1} and the  estimation policy has the structure as in \cref{thm: optimal_estimator_n_m} satisfies
\begin{equation}
	\varphi^\star=
	\left\{\begin{array}{ll}
		{\tilde{\varphi}} & {\text {if }  \mathbf{E}\big[ \|X\|^2 \textbf{1}\big(\|X\|^2 > c\big) \big]> d} \\
		{0} & {\text {otherwise}.} 
	\end{array}\right.
\end{equation}
\end{theorem}
\begin{IEEEproof} For brevity, we define $\mathcal{J}\big((\gamma_{\eta^\star,\varphi}^\star,\eta^\star),\varphi\big) \Equaldef \tilde{\mathcal{J}}\big(\varphi\big)$. Reformulating $\tilde{\mathcal{J}}\big(\varphi\big)$, we obtain
\begin{multline}
	\tilde{\mathcal{J}}\big(\varphi\big)= \mathbf{E}\Big[(1-\varphi) \|X\|^2 \textbf{1}\big((1-\varphi) \|X\|^2 \le c\big) \Big] \nonumber\\ +\mathbf{E}\Big[c\textbf{1}\big((1-\varphi) \|X\|^2 > c\big) \Big] + \varphi\Big(\mathbf{E}\Big[\|X\|^2 \Big] -d \Big) \\
	= \mathbf{E}\Big[\big((1-\varphi) \|X\|^2-c\big) \textbf{1}\big((1-\varphi) \|X\|^2 \le c\big) \Big] \nonumber\\ + c + \varphi\Big(\mathbf{E}\big[\|X\|^2 \big] -d \Big).
\end{multline}

Before computing $\nabla_{\varphi} \tilde{\mathcal{J}}\big(\varphi\big)$, we first prove that
\begin{multline} \label{eq: part_derivative}
	\mathcal{Q}(\varphi)\Equaldef \nabla_{\varphi}	\mathbf{E}\big[\big((1-\varphi) \|X\|^2-c\big) \textbf{1}\big((1-\varphi) \|X\|^2 \le c\big) \big] \\= -\mathbf{E}\Big[ \|X\|^2 \textbf{1}\big((1-\varphi) \|X\|^2 \le c\big) \Big].
\end{multline}
Representing $\mathcal{Q}(\varphi)$ in integral form yields
\begin{align}
	\mathcal{Q}(\varphi)= \nabla_{\varphi} \bigg[ \int_{-\sqrt{\frac{c}{1-\varphi}}}^{\sqrt{\frac{c}{1-\varphi}}} \int_{-\sqrt{\frac{c}{1-\varphi}-x_m^2}}^{\sqrt{\frac{c}{1-\varphi}-x_m^2}}\ldots \int_{-\sqrt{\frac{c}{1-\varphi}-\sum_{i=2}^m x_i^2}}^{\sqrt{\frac{c}{1-\varphi}-\sum_{i=2}^m x_i^2}} \nonumber \\
	\frac{(1-\varphi) \|x\|^2-c}{\sqrt{(2\pi)^m|{\Sigma}|}} \exp\left( -\frac{1}{2}{x}^\mathsf{T} {\Sigma}^{-1} {x} \right) \mathrm{d} {x}_1\ldots \mathrm{d} {x}_{m-1} \mathrm{d} {x}_m\bigg].
\end{align}
Using the Leibniz's rule $m$ times, we get
\begin{align}
	&\mathcal{Q}(\varphi) \stackrel{(a)}{=}  \int_{-\sqrt{\frac{c}{1-\varphi}}}^{\sqrt{\frac{c}{1-\varphi}}} \int_{-\sqrt{\frac{c}{1-\varphi}-x_m^2}}^{\sqrt{\frac{c}{1-\varphi}-x_m^2}}\ldots \nabla_{\varphi} \bigg[\int_{-\sqrt{\frac{c}{1-\varphi}-\sum_{i=2}^m x_i^2}}^{\sqrt{\frac{c}{1-\varphi}-\sum_{i=2}^m x_i^2}} \nonumber \\
	&\frac{(1-\varphi) \|x\|^2-c}{\sqrt{(2\pi)^m|{\Sigma}|}} \exp\left( -\frac{1}{2}{x}^\mathsf{T} {\Sigma}^{-1} {x} \right) \mathrm{d} {x}_1 \bigg] \ldots \mathrm{d} {x}_{m-1} \mathrm{d} {x}_m\\
	&\stackrel{(b)}{=} - \int_{-\sqrt{\frac{c}{1-\varphi}}}^{\sqrt{\frac{c}{1-\varphi}}} \int_{-\sqrt{\frac{c}{1-\varphi}-x_m^2}}^{\sqrt{\frac{c}{1-\varphi}-x_m^2}}\ldots \int_{-\sqrt{\frac{c}{1-\varphi}-\sum_{i=2}^m x_i^2}}^{\sqrt{\frac{c}{1-\varphi}-\sum_{i=2}^m x_i^2}} \nonumber \\
	&\frac{\|x\|^2}{\sqrt{(2\pi)^m|{\Sigma}|}} \exp\left( -\frac{1}{2}{x}^\mathsf{T} {\Sigma}^{-1} {x} \right) \mathrm{d} {x}_1  \ldots \mathrm{d} {x}_{m-1} \mathrm{d} {x}_m\\
	&=-\mathbf{E}\Big[ \|X\|^2 \textbf{1}\big((1-\varphi)\|X\|^2 \le c\big) \Big].
\end{align}
where $(a)$ uses the fact that the integrands are zero at the integral limits and $(b)$ uses
{\small
\begin{align}
	&\nabla_{\varphi} \bigg[\int_{-\sqrt{\frac{c}{1-\varphi}-\sum_{i=2}^m x_i^2}}^{\sqrt{\frac{c}{1-\varphi}-\sum_{i=2}^m x_i^2}} \frac{(1-\varphi) \|x\|^2-c}{\sqrt{(2\pi)^m|{\Sigma}|}}  \exp\left( -\frac{1}{2}{x}^\mathsf{T} {\Sigma}^{-1} {x} \right) \mathrm{d} {x}_1 \bigg] \nonumber \\
	&= \int_{-\sqrt{\frac{c}{1-\varphi}-\sum_{i=2}^m x_i^2}}^{\sqrt{\frac{c}{1-\varphi}-\sum_{i=2}^m x_i^2}} \frac{- \|x\|^2}{\sqrt{(2\pi)^m|{\Sigma}|}}  \exp\left( -\frac{1}{2}{x}^\mathsf{T} {\Sigma}^{-1} {x} \right) \mathrm{d} {x}_1.
\end{align}}

Taking the derivative of $\tilde{\mathcal{J}} \big(\varphi\big)$ with respect to $\varphi$ and using \cref{eq: part_derivative}, we have
\begin{align}
	\nabla_{\varphi} \tilde{\mathcal{J}} \big(\varphi\big)=\ & \mathcal{Q}(\varphi) +\mathbf{E}\big[\|X\|^2 \big] -d \nonumber\\
	=\ & \mathbf{E}\Big[ \|X\|^2 \textbf{1}\big((1-\varphi) \|X\|^2 > c\big) \Big] -d.
\end{align}

Then we use the same technique as the proof of \cref{thm: optimal_jam_prob} to complete the proof.
\end{IEEEproof}

}


Once the result is established for diagonal covariance matrices, it is simple to obtain the result for general covariance matrices such that $\Sigma$ is symmetric positive definite. Since $\Sigma$ admits an eigendecomposition such that $\Sigma = W^\mathsf{T}\Lambda W$, where $\Lambda$ is a diagonal matrix and $W^\mathsf{T}W=I$. Thus, equipped with \cref{thm: optimal_jam_prob_m}, upon observing $x_i$, each agent computes $\tilde{x}_i=Wx_i$, uses its transmission policy designed by the coordinator assuming the covariance matrix is $\Lambda$. Notice that the transmission decision is computed using $\tilde{x}_i$ and not $x_i$. However, when a packet is transmitted, agent $i$ sends the original observation $x_i$. This scheme preserves the saddle point equilibrium property, and therefore, our results hold in full generality for multivariate Gaussian observations.

\section{Proof of \cref{thm:jamming_Lagrange}} \label{appendix:large_scale}

\begin{IEEEproof}
	We first obtain the derivatives of $\tilde{\tilde{L}}(\varphi, \lambda)$ with respect to $\varphi$ and $\lambda$, which are given by
    \begin{equation} \label{eq: L_d_lambda}
		{\nabla_\lambda}\tilde{\tilde{L}}(\varphi, \lambda)
		=  2\int_{\sqrt{(c+\lambda)/(1-\varphi)}}^{+\infty} f(x)\mathrm{d} x - \bar{\kappa}
	\end{equation}
	and
	\begin{equation} \label{eq: L_d_phi}
		{\nabla_\varphi}\tilde{\tilde{L}}(\varphi, \lambda)
		=  2\int_{\sqrt{(c+\lambda)/(1-\varphi)}}^{+\infty} x^2f(x)\mathrm{d} x -d.
	\end{equation}
	
	Then, notice that ${\nabla_\lambda}\tilde{\tilde{L}}(\varphi, \lambda)$ is a monotonic decreasing function in $\lambda$ and that
	\begin{align}
		{\nabla_\lambda}\tilde{\tilde{L}}(\varphi, 0)
		&=  2\int_{\sqrt{c/(1-\varphi)}}^{+\infty} f(x)\mathrm{d} x - \bar{\kappa}\\
		\lim_{\lambda \uparrow +\infty} {\nabla_\lambda}\tilde{\tilde{L}}(\varphi, \lambda)
		&=  - \bar{\kappa}.
	\end{align}
	Moreover, ${\nabla_\varphi}\tilde{\tilde{L}}(\varphi, \lambda)$ is a monotonic decreasing function in $\varphi$ such that 
	\begin{align}
		{\nabla_\varphi}\tilde{\tilde{L}}(0, \lambda)
		&=  2\int_{\sqrt{c+\lambda}}^{+\infty} x^2f(x)\mathrm{d} x -d \\
		\lim_{\varphi\uparrow 1} {\nabla_\varphi}\tilde{\tilde{L}}(\varphi, \lambda)
		&=  -d.
	\end{align}

Based on the aforementioned facts, we can establish the optimal jamming probability and Lagrangian dual variable in different cases:
	1) When $\int_{\sqrt{c}}^{+\infty} f(x) \mathrm{d} x <\bar{\kappa}/2$ and $\int_{\sqrt{c}}^{+\infty} x^2 f(x)\mathrm{d} x <d/2$, we have
	\begin{equation}
		{\nabla_\lambda}\tilde{\tilde{L}}(\varphi, 0)< 0 \text{~~and~~} {\nabla_\varphi}\tilde{\tilde{L}}(0, \lambda)< 0.
	\end{equation}
	Since the function $\tilde{\tilde{L}}(\varphi, \lambda)$ is decreasing in $\lambda$ for fixed $\varphi$ and decreasing in $\varphi$ for fixed $\lambda$, the optimal $\lambda^\star$ and $\varphi^\star$ must be
	\begin{equation}
		\lambda^\star=0 \text{~~and~~} \varphi^\star=0.
	\end{equation}

	2) When $\int_{\sqrt{c}}^{+\infty} f(x) \mathrm{d} x <\bar{\kappa}/2$ and $\int_{\sqrt{c}}^{+\infty} x^2 f(x) \mathrm{d} x \ge d/2$, we have
	\begin{equation}
	{\nabla_\lambda}\tilde{\tilde{L}}(\varphi, 0)< 0.
	\end{equation}
	Since $\tilde{\tilde{L}}(\varphi, \lambda)$ is monotonic decreasing in $\lambda$, 
	 the optimal  $\lambda^\star$ must be
	\begin{equation}
		\lambda^\star=0.
	\end{equation}
	On the otherhand, ${\nabla_\varphi}\tilde{\tilde{L}}(0, \lambda^\star) \ge 0$ and  ${\nabla_\varphi}\tilde{\tilde{L}}(1, \lambda^\star) <0$. Therefore, due to the mononicity of $\tilde{\tilde{L}}(\varphi,\lambda)$  the optimal $\varphi$ exists, is unique and is given by
	\begin{equation}
	\varphi^\star=1-\frac{c}{l_{\varphi}(d)}.
	\end{equation}

	3) When $\int_{\sqrt{c}}^{+\infty} f(x) \mathrm{d} x \ge \bar{\kappa}/2$ and $\int_{\sqrt{c}}^{+\infty} x^2 f(x) \mathrm{d} x < d/2$, we have
	\begin{equation}
		{\nabla_\varphi}\tilde{\tilde{L}}(0, \lambda)< 0,
	\end{equation}
		Since $\tilde{\tilde{L}}(\varphi, \lambda)$ is monotonic decreasing in $\varphi$, the optimal  $\varphi^\star$ must be
	\begin{equation}
		\varphi^\star=0.
	\end{equation}
	On the otherhand, ${\nabla_\lambda}\tilde{\tilde{L}}(\varphi^\star, 0) \ge 0$ and ${\nabla_\lambda}\tilde{\tilde{L}}(\varphi, +\infty)<0$. Therefore, due to the mononicity of $\tilde{\tilde{L}}(\varphi,\lambda)$,  the optimal $\lambda$ exists, is unique and is given by
	\begin{equation}
	\lambda^\star=l_{\lambda}(\bar{\kappa})-c.
	\end{equation}

	
	4) When $\int_{\sqrt{c}}^{+\infty} f(x) \mathrm{d} x \ge \bar{\kappa}/2$ and $\int_{\sqrt{c}}^{+\infty} x^2 f(x) \mathrm{d} x \ge d/2$,

		\noindent (a)  When $l_{\lambda}=l_{\varphi}$, then all pairs $(\lambda^\star,\varphi^\star)$ satisfying
		\begin{align}
			\frac{c+\lambda^\star}{1-\varphi^\star}=l_{\lambda}
		\end{align}
		are optimal.\\
		(b)  When $l_{\lambda}(\bar{\kappa})>l_{\varphi}(d)$, exactly one of the two values ($\lambda^\star$,$\varphi^\star$) is nonzero. This is because the pair $(0,0)$ would violate the assumption that $\int_{\sqrt{c}}^{+\infty} f(x) \mathrm{d} x \ge \bar{\kappa}/2$ and $\int_{\sqrt{c}}^{+\infty} x^2 f(x) \mathrm{d} x \ge d/2$. The pair $\big(l_{\lambda}(\bar{\kappa})-c,1-c/l_{\varphi}(d)\big)$ cannot be optimal since the first order optimality conditions would contradict $l_{\lambda}(\bar{\kappa})>l_{\varphi}(d)$. Therefore, the optimal pair is either $\big(0,1-c/l_{\varphi}(d)\big)$ or $(l_{\lambda}-c,0)$. However, the pair $\big(0,1-c/l_{\varphi}(d)\big)$ cannot be optimal since $\lambda^\star=0$ implies that ${\nabla_\lambda}\tilde{L}(\varphi^\star, 0)=2\int_{\sqrt{l_{\varphi}}}^{+\infty} f(x) \mathrm{d} x-\bar{\kappa}<0$, which violates the assumption that $2\int_{\sqrt{l_{\varphi}}}^{+\infty} f(x) \mathrm{d} x-\bar{\kappa}> 2\int_{\sqrt{l_{\lambda}}}^{+\infty} f(x) \mathrm{d} x-\bar{\kappa} =0$. Finally, the only pair that satisfies the first-order optimality conditions and do not lead to contradictions is
		\begin{equation}
			\lambda^\star=l_{\lambda}(\bar{\kappa})-c \text{~~and~~} \varphi^\star=0.
		\end{equation}
		(c)  Similar to the previous case, when $l_{\varphi}(d)>l_{\lambda}(\bar{\kappa})$, we obtain  
		\begin{equation}
			\lambda^\star=0 \text{~~and~~} \varphi^\star=1-\frac{c}{l_{\varphi}(d)}.
		\end{equation}
\end{IEEEproof}

\vspace{-0.5in}

\begin{IEEEbiography}[{\includegraphics[width=1in,height=1.25in,clip,keepaspectratio]{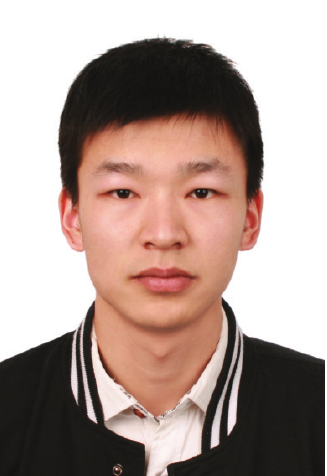}}]{Xu Zhang} is a Postdoctoral Researcher at the State Key Laboratory of Scientific and Engineering Computing, Academy of Mathematics and Systems Science, Chinese Academy of Sciences. He received his B.S. degree and Ph.D. degree in Electronics Engineering from the School of Information and Electronics, Beijing Institute of Technology, Beijing, China, in 2015 and 2021. He was a visiting student in the Ming Hsieh Department of Electrical Engineering at the University of Southern California from 2018 to 2019. His research interests include federated learning, distributed optimization, compressed sensing, and inverse problems.
\end{IEEEbiography}	

\vspace{-0.5in}
\begin{IEEEbiography}[{\includegraphics[width=1in,height=1.25in,clip,keepaspectratio]{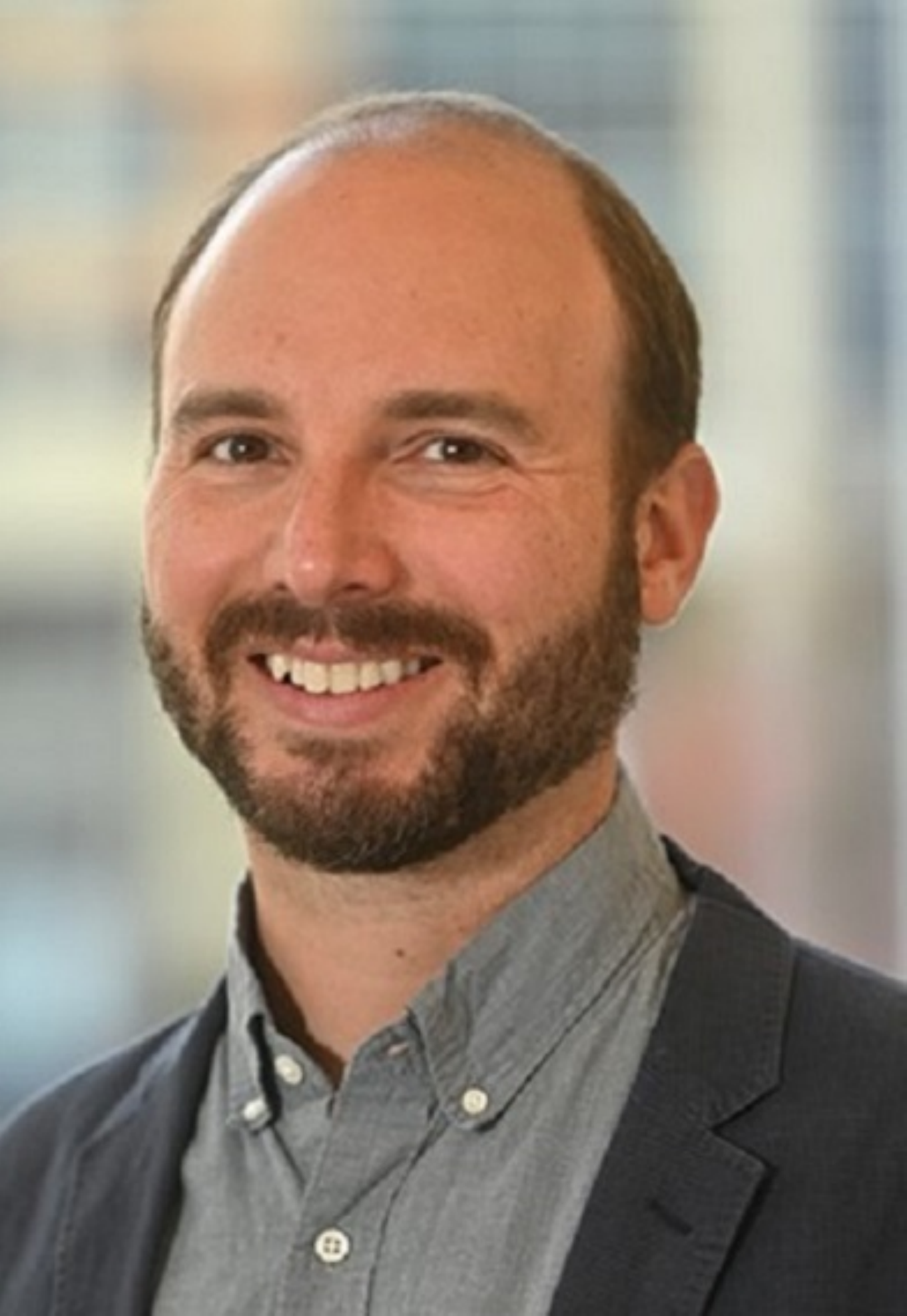}}]{Marcos M. Vasconcelos} is an Assistant Professor with the Department of Electrical Engineering at the FAMU-FSU College of Engineering, Florida State University. He received his Ph.D. from the University of Maryland, College Park, in 2016. He was a Research Assistant Professor at the Commonwealth Cyber Initiative and the Bradley Department of Electrical and Computer Engineering at Virginia Tech from 2021 to 2022. From 2016 to 2020, he was a Postdoctoral Research Associate in the Ming Hsieh Department of Electrical Engineering at the University of Southern California. His research interests include networked control and estimation, multi-scale robotic networks, game theory, multi-agent optimization, distributed machine learning, and systems biology.
\end{IEEEbiography}

\end{document}